\documentclass[useAMS,fleqn,usenatbib]{mnras_anon}
\usepackage{graphicx}
\usepackage{hyperref}
\usepackage{bm}
\usepackage{float}

\mathchardef\mhyphen="2D

\newcommand{\origami}{{\scshape origami}}

\newcommand{\bA}{\bm{A}}
\newcommand{\bB}{\bm{B}}
\newcommand{\ba}{\bm{a}}
\newcommand{\bb}{\bm{b}}

\newcommand{\bq}{\bm{q}}
\newcommand{\bp}{\bm{p}}

\newcommand{\bx}{\bm{x}}

\newcommand{\bv}{\bm{v}}

\newcommand{\hmpc}{\,$h^{-1}$\,Mpc}

\newcommand{\hmpcnosp}{$h^{-1}$\,Mpc}

\newcommand{\bnabla}{\mbox{\boldmath $\nabla$}}
\newcommand{\bPsi}{\mbox{\boldmath $\Psi$}}

\newcommand{\citepp}{\citep}
\newcommand{\citett}{\citet}
\newcommand{\hrefurl}[1]{\href{#1}{#1}}

\chardef\til=`\~

\pubyear{2017}

\begin{document}

\title[The cosmic spiderweb]{The cosmic spiderweb: equivalence of cosmic, architectural, and origami tessellations}

\author[Mark C.\ Neyrinck et al.]
{Mark Neyrinck$^1$, Johan Hidding$^{2,3}$, Marina Konstantatou$^4$, Rien van de Weygaert$^2$\\
$^1$Institute for Computational Cosmology, Department of Physics, Durham University, Durham, DH1 3LE, UK\\
$^2$Kapteyn Astronomical Institute, Univ.\ Groningen, P.O.\ Box 800, Groningen, the Netherlands\\
$^3$Netherlands eScience Center, Science Park 140, 1098 XG, Amsterdam, the Netherlands\\
$^4$Department of Engineering, University of Cambridge, Cambridge, CB2 1PZ, UK}


\maketitle

\begin{abstract}
For over twenty years, the term `cosmic web' has guided our understanding of the large-scale arrangement of matter in the cosmos, accurately evoking the concept of a network of galaxies linked by filaments. But the physical correspondence between the cosmic web and structural-engineering or textile `spiderwebs' is even deeper than previously known, and extends to origami tessellations as well. Here we explain that in a good structure-formation approximation known as the adhesion model, threads of the cosmic web form a spiderweb, i.e.\ can be strung up to be entirely in tension. The correspondence is exact if nodes sampling voids are included, and if structure is excluded within collapsed regions (walls, filaments and haloes), where dark-matter multistreaming and baryonic physics affect the structure. We also suggest how concepts arising from this link might be used to test cosmological models: for example, to test for large-scale anisotropy and rotational flows in the cosmos.
\end{abstract}


\begin {keywords}
  large-scale structure of Universe -- cosmology: theory -- structural engineering -- reciprocal diagrams -- graphic statics -- computational geometry
\end {keywords}

\section{Introduction}
Textile concepts in mythological cosmology go back to antiquity, e.g.\ to the Fates spinning a tapestry representing destiny, and the planets riding on spheres rotating about a cosmic spindle\footnote{Spindle in the sense of a rod for spinning wool into yarn, mentioned in Plato's {\it Republic} \citepp[discussed by][]{James1995}.}. In modern physical cosmology as well, the textile concept of a cosmic web is important, the term introduced and popularized in the paper `How filaments of galaxies are woven into the cosmic web' \citepp{BondEtal1996}. `Web' was first used in this context even before, in the 1980's \citepp{Shandarin1983, KlypinShandarin1983}.

%

The cosmic web has inspired many textile artistic representations, reviewed by \citett{DiemerFacio2017}, who also detail new techniques for `tactilization' of the cosmic web. The cosmic web has only recently been accurately mapped, but it resembles some more familiar natural structures. These bear similarities to human-designed structures both because of human inspiration, and because the same mathematical and engineering principles apply both to nature and human design \citepp{ArslanSorguc2004}.

The cosmic web and its resemblance to a spiderweb has inspired several room-sized installations by Tom\'{a}s Saraceno\footnote{See e.g.\ {\it How to Entangle the Universe in a Spider Web}, {\it 14 Billions} and {\it Galaxies forming along filaments, like droplets along the strands of a spiders web}, at \hrefurl{http://www.tomassaraceno.com}.}, described by \citett{Ball2017}.  Referencing Saraceno's work, \citett{Livio2012} mentions `the visual (although clearly not physical) similarity between spider webs and the cosmic web.' This paper responds to `clearly not physical'; we explain the physical similarity between spider and cosmic webs.

This similarity is through a geometric concept of a `spiderweb' used in architecture and engineering, related to generalized Voronoi and Delaunay tessellations \citepp[e.g.][]{OkabeEtal2008,AurenhammerEtal2013}. Such tessellations are already crucial concepts and tools in cosmology \citepp[e.g.][]{VdwEtal2009,NeyrinckShandarin2012}. The Voronoi foam concept \citepp{IckeVdw1987,VdwIcke1989,Vdw2007} has also been instrumental in shaping our understanding of the large-scale arrangement of matter and galaxies in the universe. Indeed, the arrangement of matter on large scales behaves in many respects like a cellular system \citepp{AragonCalvo2014}. Tessellation concepts have been used for some time in cosmological data analysis, as well \citepp[e.g.][]{BernardeauVdw1996,SchaapVdw2000,NeyrinckEtal2005,vdwSchaap2009}. For more information about the cosmic web, see recent proceedings of `The Zeldovich Universe: Genesis and Growth of the Cosmic Web' symposium \citepp{VdwShandarinSaar2016} and a comparison of ways of classifying parts of the cosmic web \citepp{LibeskindEtal2017}.

In the real Universe, filaments of galaxies between clusters of galaxies have been observed for some time \citep[e.g.][]{deLapparentEtal1986}. Observing the cosmic web of filaments of dark and ordinary baryonic matter between galaxies is much more difficult, but has started to happen. With weak gravitational lensing observations, large filaments of matter have been detected individually \citep{JauzacEtal2012}; smaller filaments have been detected by stacking filament signals together \citep{ClampittEtal2016}. Very recently, filaments in ionized gas have been detected \citep{TanimuraEtal2017,deGraaffEtal2017}, by stacking their Sunyaev-Zeldovich effect, a spectral distortion that ionized gas imparts onto the primordial cosmic microwave background radiation. One reason these observations are important is that they support the standard picture that much (tens of percent) of the baryonic matter in the Universe resides not in galaxies themselves, but in the cosmic web of filaments between galaxies.

In this paper, first we will discuss the spiderweb concept and how it applies to large-scale structure, in 1D, 2D, and 3D. Then, in \S \ref{sec:applications}, we suggest some applications of these ideas beyond the understanding of structure formation itself.


\section{Geometry of spiderwebs}
First we will define the geometric concept of a {\it spiderweb} \citepp[e.g.][]{WhiteleyEtal2013}, a particular type of spatial graph. Conceptually, a spiderweb is a spatial (i.e.\ with nodes at specified positions) graph that can be strung up in equilibrium such that all of its strings are in tension. It need not look like the common conception of a biological spiderweb, i.e.\ a 2D pattern of concentric polygons supported by radial strands. We will discuss the various concepts in some detail first in 2D, and then turn to 3D. Graphic statics, a visual design and analysis method based on the idea of reciprocity between form and force diagrams, developed after Maxwell's work in the 19th century \citepp[e.g.][]{Kurrer2008}. It tailed off in popularity in the early 20th century, but has recently experienced a resurgence of interest, especially the study of global static equilibrium for both 2D and 3D structures \citepp[e.g.][]{McRobie2016,KonstantatouMcRobie2016,McRobie2017,BlockEtal2016,AkbarzadehEtal2016}.

The key mathematical property that characterizes a spiderweb is a perpendicularity property. Consider a graph with positioned nodes, and straight edges between them. Suppose that the graph is planar (i.e.\ with non-crossing edges), and so the graph tessellates the plane into polygonal cells. Further suppose a set of {\it generators} exists, one per cell, such that the line connecting the generators of each pair of bordering cells is perpendicular to the edge that forms their border. The network of edges connecting bordering cells is called the {\it dual}, and if those edges are perpendicular to cells' borders, it is a {\it reciprocal} dual. The original graph is a {\it spiderweb} if the edges of the dual tessellation do not cross each other.

\begin{figure}
    \begin{center}
    \includegraphics[width=\columnwidth]{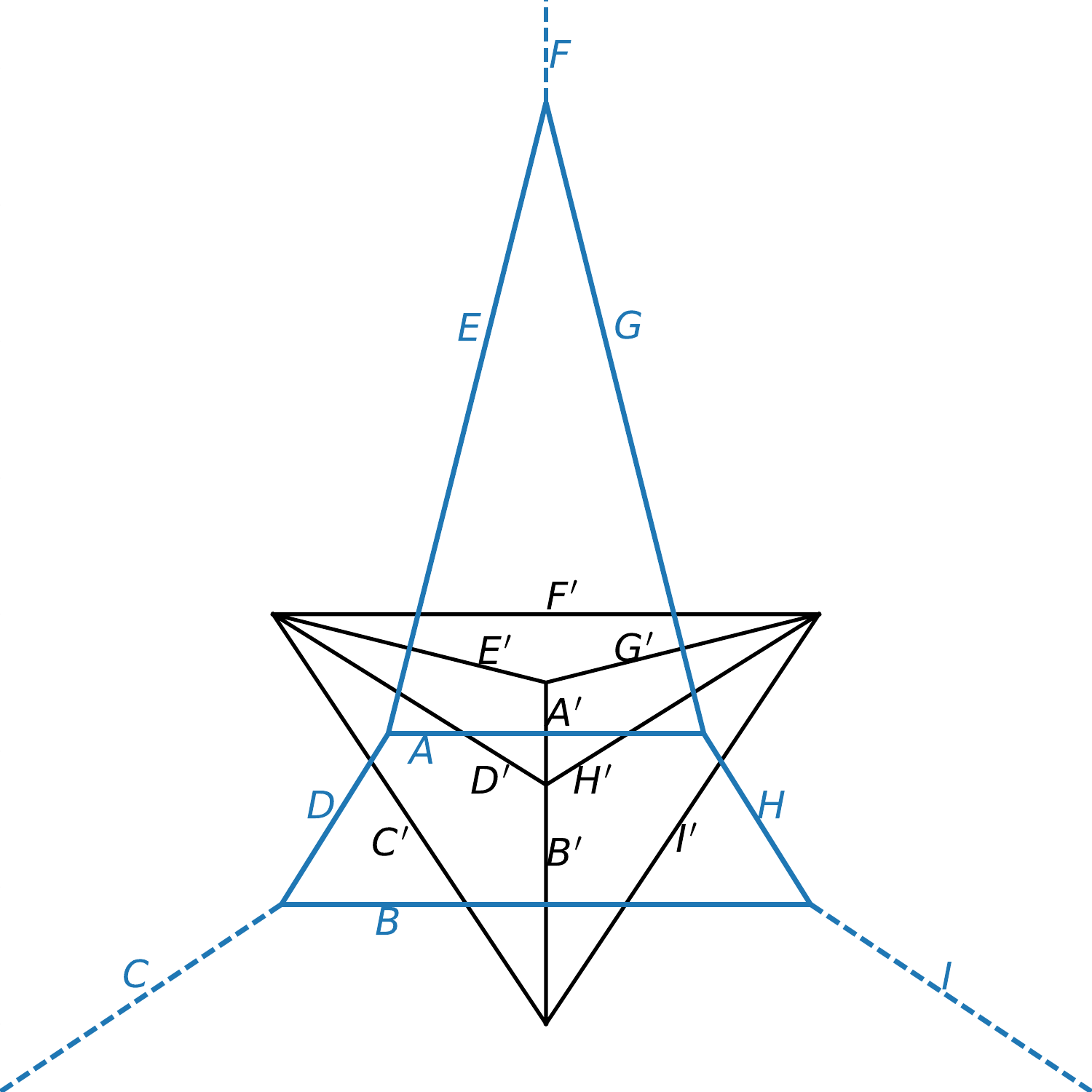}
  \end{center}  
  \caption{A spiderweb form diagram (blue) resembling the Eiffel Tower, and the corresponding force diagram (black). Letters label perpendicular pairs of form (unprimed) and force (primed) edges. Some perpendicular segment pairs would only actually intersect if extended. Dashed edges lead to external supports.}
  \label{fig:eiffeltower}
\end{figure}

James Clerk Maxwell, better known in physics for uniting electromagnetism, conceived of reciprocal duals to analyze and design pin-jointed trusses in structural engineering \citepp{Maxwell1864,Maxwell1867}. A {\it form diagram} is just a map of the structural members in a truss, containing nodes and edges between them. A {\it force diagram} also has one edge per beam, but the length of each `form' edge is proportional to the internal force (if positive, compression; if negative, tension) in that structural member. Maxwell showed that if the network is in equilibrium, a closed force diagram can be constructed such that the form and force networks are reciprocals of each other. Fig.\ \ref{fig:eiffeltower} shows an example of a spiderweb. The reciprocal-dual force diagram appears in black; the form diagram is a spiderweb because all force polygons are closed and fit together without crossing any edges. 

To gain some intuition, consider all forces acting on a node of the form diagram. They can be represented by a {\it force polygon} around it, each side perpendicular to an edge coming off the node, of length proportional to the tension. The condition of the node's forces being in balance is equivalent to the force polygon being closed. Now consider adjacent nodes, with their own force polygons. The tension and direction of each edge will be the same at both ends, so the sides of the force polygons at both ends will be the same as well; they may be neatly fitted together without gaps. The ensemble of all such polygons constitutes the force diagram.

The force and form diagrams in Fig.\ \ref{fig:eiffeltower} are Voronoi and Delaunay tessellations, their generating points lying at the vertices of the Delaunay force diagram. Each blue Voronoi cell in the form diagram is the set of points in space closer to its generator point than to any other. The black, force diagram is the Delaunay tessellation joining generators. Voronoi and Delaunay tessellations are reciprocal duals, and are always spiderwebs; they have non-crossing, convex polygons, with edges of one perpendicular to edges of the other. 

Fig.\ \ref{fig:eiffeltower} is supposed to resemble the Eiffel Tower, designed principally by Maurice Koechlin using graphics statics. Koechlin apparently wrote only one article not strictly about engineering, and that happens to be about how spiders spin their webs, mentioning some structural issues \citepp{Koechlin1905}. See \citett{FivetEtal2015} for further information about Koechlin and his publications.

The set of spiderwebs is more general than the set of Voronoi diagrams, but only a bit. Each edge of the Voronoi diagram can slide perpendicular to its dual edge (the Delaunay edge joining generators). This edge can be slid in constructing the tessellation by adding a different constant to the distance functions used to decide which points are closest to the generators forming that edge.

In symbols, the cell $V_{\bq}$ around the generator at position $\bq$ is
\begin{equation}
V_{\bq}=\left\{\bx\in \mathcal{E}~{\rm s.t.}~ |\bx-\bq|^2 + z_q^2 \le |\bx-\bp|^2 + z_p^2,~\forall \bp\in \mathcal{L}\right\},
\label{eqn:secvoronoi}
\end{equation}
where $\mathcal{E}$ is the space being tessellated, $\mathcal{L}$ is the set of generator points, and $z_q^2$ and $z_p^2$ are constants, possibly different for each generator. If all constants $z$ are equal, both sides of the inequality reduce to the usual distance functions, giving a usual Voronoi diagram. If they differ, the tessellation generalizes to a {\it sectional-Voronoi diagram} (also known as a `power diagram'), generally not a Voronoi diagram \citep{ChiuEtal1996}. The reason for the word `sectional' is that the tessellation can be seen as a cross-section through a Voronoi diagram in a higher dimension, with each $z_q$ interpreted as a distance from the cross-section. In 2D, the set of spiderweb networks is exactly the set of sectional-Voronoi diagrams \citepp{AshBolker1986,WhiteleyEtal2013}, and 3D sectional-Voronoi diagrams are guaranteed to be spiderwebs, as well.

\subsection{Non-spiderwebs}
What kinds of spatial graphs are non-spiderweb? One clear sign that a spatial graph is non-spiderweb is if any of its polygons are non-convex. Suppose there is a non-convex polygon, so there is at least one node of the polygon whose interior angle exceeds 180\degr. That means that all of the threads pulling it are in the same half-plane, with no balancing tension in the other directions, so this node cannot be in force equilibrium.

\begin{figure}
    \begin{center}
    \includegraphics[width=0.7\columnwidth]{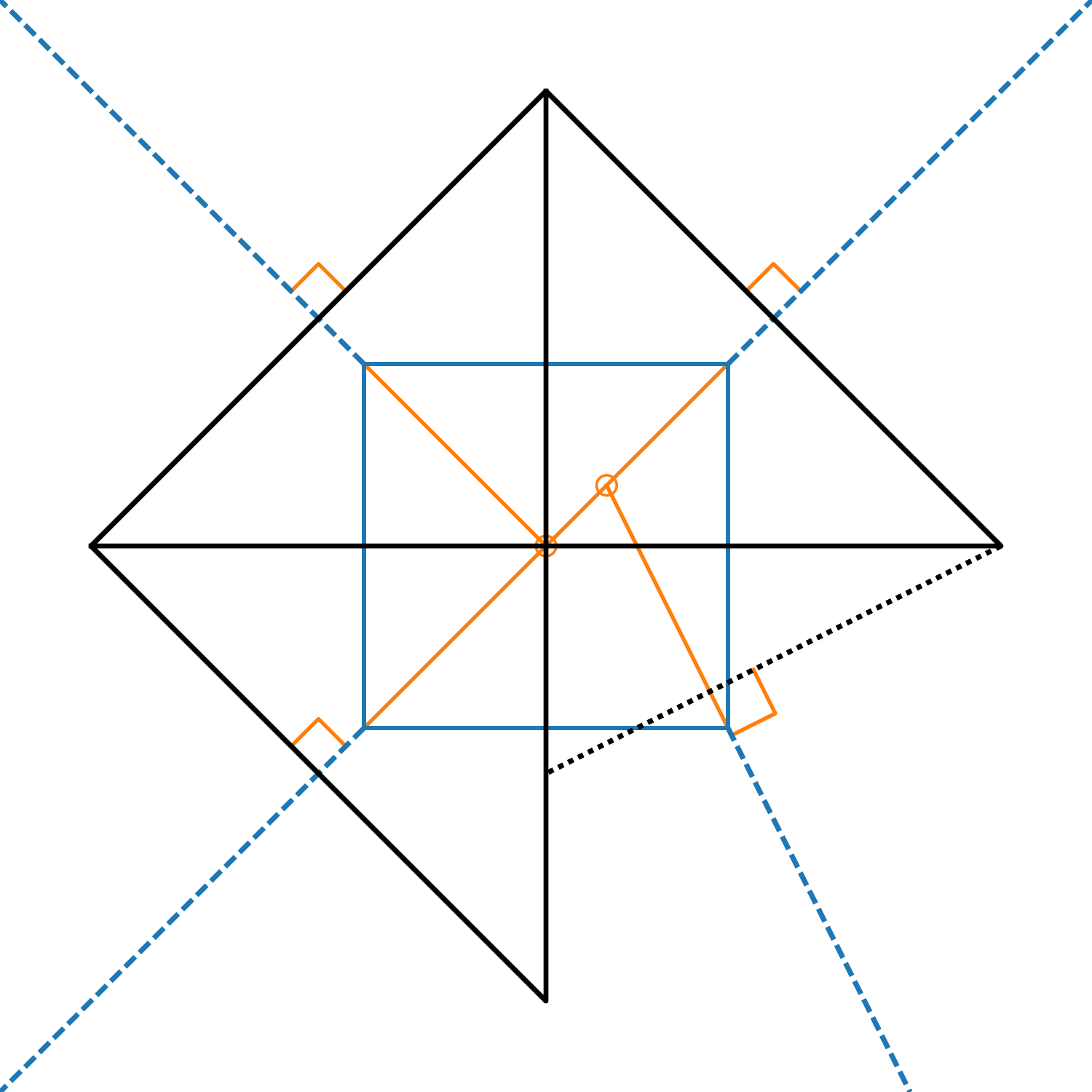}
  \end{center}  
  \caption{{\it Blue:} A spatial graph that is not a spiderweb. {\it Black:} An attempted force diagram. It cannot close with the bottom-right member askew, so it would not be in equilibrium if strung up in tension. {\it Orange}: continuations of the external members, that do not meet in a point.}
  \label{fig:nonspiderweb}
\end{figure}

Fig. \ref{fig:nonspiderweb} shows another kind of non-spiderweb. Here, members of the blue form diagram outline a square, supported by external members (dashed blue lines) at each corner. Three of these point out 45\degr-diagonally; their force polygons are 45\degr\ right triangles. But the lower-right external member has a different angle; this is a problem because this means the node's force polygon must be differently-shaped, like the bottom-left triangle. There is no way to scale this triangle to join both vertices to the adjacent triangles, so this is not even a reciprocal dual.

Physical intuition might be of some help to understand this. Imagine stringing up this pattern in tension, such that all strings pointed radially out from the center of the square. Now imagine changing the angle of one of the strings. After the pattern equilibrates, it would be distorted.

Similarly, if the angle of one of the external members of Fig.\ \ref{fig:eiffeltower} were changed, the structure would not be in equilibrium. This is related to a so-called Desargues configuration, named after a founder of projective geometry, a subject closely related to graphic statics \citepp[e.g.][]{CrapoWhiteley1982,KonstantatouMcRobie2016}. If external members extend from vertices of a triangle in equilibrium (e.g. Fig.\ \ref{fig:eiffeltower}, without the central member $A$), those members' vectors, extended inside the triangle, must meet in a point. In perspective geometry, this is called the center of perspectivity, the vanishing point of parallel rays at infinity. In Fig.\ \ref{fig:eiffeltower}, lines $C$, $F$ and $I$ meet in such a center of perspectivity, as is guaranteed by $C^\prime$, $F^\prime$ and $I^\prime$ forming a closed triangle.

For a triangle, given two external member directions, the third is fixed also to point to where they would intersect. For polygons with more sides, a simple way to force spiderwebness is still for external member vectors to meet in a single point. (Perhaps this is a reason why idealized biological spiderwebs have concentric polygons joined by radial strands that intersect in the center.) But there are equilibrium spiderweb configurations without members meeting in a single point as well; for an $N$-sided polygon, the closing of the form diagram only fixes one remaining member direction given all $N-1$ others.

\section{The adhesion model and spiderwebs}
Remarkably, the accurate {\it adhesion model} of large-scale structure produces a cosmic web that is exactly a sectional-Voronoi diagram, as well.
 
The Zeldovich approximation \citepp[][ZA]{Zeldovich1970} is the usual starting point for the adhesion model, although any model for a displacement potential (defined below) can be used. The ZA is a first-order perturbation theory, perturbing particle displacements away from an initial uniform grid, and already describes the morphology of the cosmic web remarkably well \citepp[e.g.][]{ColesEtal1993}. But it grows notably inaccurate after particles cross in their ballistic trajectories. Crossing is allowed physically because the (dark) matter is assumed to be collisionless, but $N$-body simulations of full gravity show that gravity itself is enough to keep these dark-matter structures compact once they form. `Collapse' is our term for the structure forming, i.e.\ for particle trajectories crossing, forming a {\it multistream} region.

The adhesion model \citepp{GurbatovSaichev1984,KofmanEtal1990,GurbatovEtal2012} eliminates the ZA over-crossing problem with a mechanism that sticks trajectories together when they cross. In the adhesion model, a viscosity $\nu$ is introduced formally into the equation of motion (resulting in a differential equation called Burgers' equation), and then the limit is taken $\nu\to 0$. It can then be solved elegantly, by a few methods, including a Laplace transform, and the following convex hull construction \citepp{VergassolaEtal1994}.

Let $\bPsi(\bq)\equiv \bx-\bq$ denote the {\it displacement} field, the displacement between the final ({\it Eulerian}) position $\bx$ and initial ({\it Lagrangian}) position ($\bq$) of a particle. These labels based on mathematicians' names refer to coordinate systems in fluid dynamics. All distances here are {\it comoving}, meaning that the expansion of the Universe is scaled out. The initial velocity field $\bv(\bq)$ after inflation is usually assumed to have zero vorticity \citep[e.g.][]{Peebles1980}, any primordial rotational modes having been damped away through inflationary expansion. In the ZA, $\bPsi\propto\bv$, so $\bPsi$ is also curl-free. In full gravity, $\bPsi$ does have a curl component in collapsed regions, but in uncollapsed regions, it seems to be tiny compared to the divergence \citepp{Chan2014,Neyrinck2016}. So here we assume that $\bPsi(\bq)=-\bnabla_{\bq}\Phi$, for a displacement potential $\Phi(\bq)$.

As discussed in the above adhesion-model papers, the mapping between $\bq$ and $\bx$ for a particle is given by the implicit equation
\begin{equation}
\Phi(\bx) = \max_q\left[\Phi(\bq)-\frac{1}{2}|\bx-\bq|^2\right].
\label{adhesionEuler}
\end{equation}
Here, $\Phi(\bx)$ is the same Lagrangian potential $\Phi$ as in $\Phi(\bq)$, but evaluated at the Eulerian position $\bx$.

For a 2D problem, one way to think about this solution is to slide an upward-opening paraboloid with equation $z=|\bx-\bq|^2/2$ around on a surface with height $z=\Phi(\bq)$, and mapping any $\bq$ patches that touch the paraboloid to the position of its minimum point, $\bx$. If the paraboloid touches more than one point, the entire polygon or polyhedron of Lagrangian space joining those points adheres together, and is placed at $\bx$. \citett{KofmanEtal1990} nicely explain this process. \citett{VergassolaEtal1994} go on to discuss how this leads to a convex-hull algorithm: raise points sampling the space in a new spatial dimension according to their displacement potential, and shrink-wrap this surface with a convex hull. Uncollapsed (defined below) regions will be on this convex hull, while collapsed regions will be inside, not touching it.

Now, the key result linking spiderwebs to the cosmic web: this convex-hull operation is equivalent to constructing a sectional-Voronoi diagram  \citett{HiddingEtal2012,HiddingEtal2016,HiddingEtal2018,Hidding2018}. This is related to the popular convex-hull method of computing a Voronoi tessellation \citepp{Brown1979}. For the following figures, we used a Python code that we provide \citepp{Hidding2017zenodo}, which uses a native Python wrapper of the Qhull \citepp{BarberEtal1996} convex hull algorithm to compute the sectional-Voronoi diagram.

\begin{figure*}
 \begin{minipage}{175mm}
    \begin{center}
    \includegraphics[width=0.8\columnwidth]{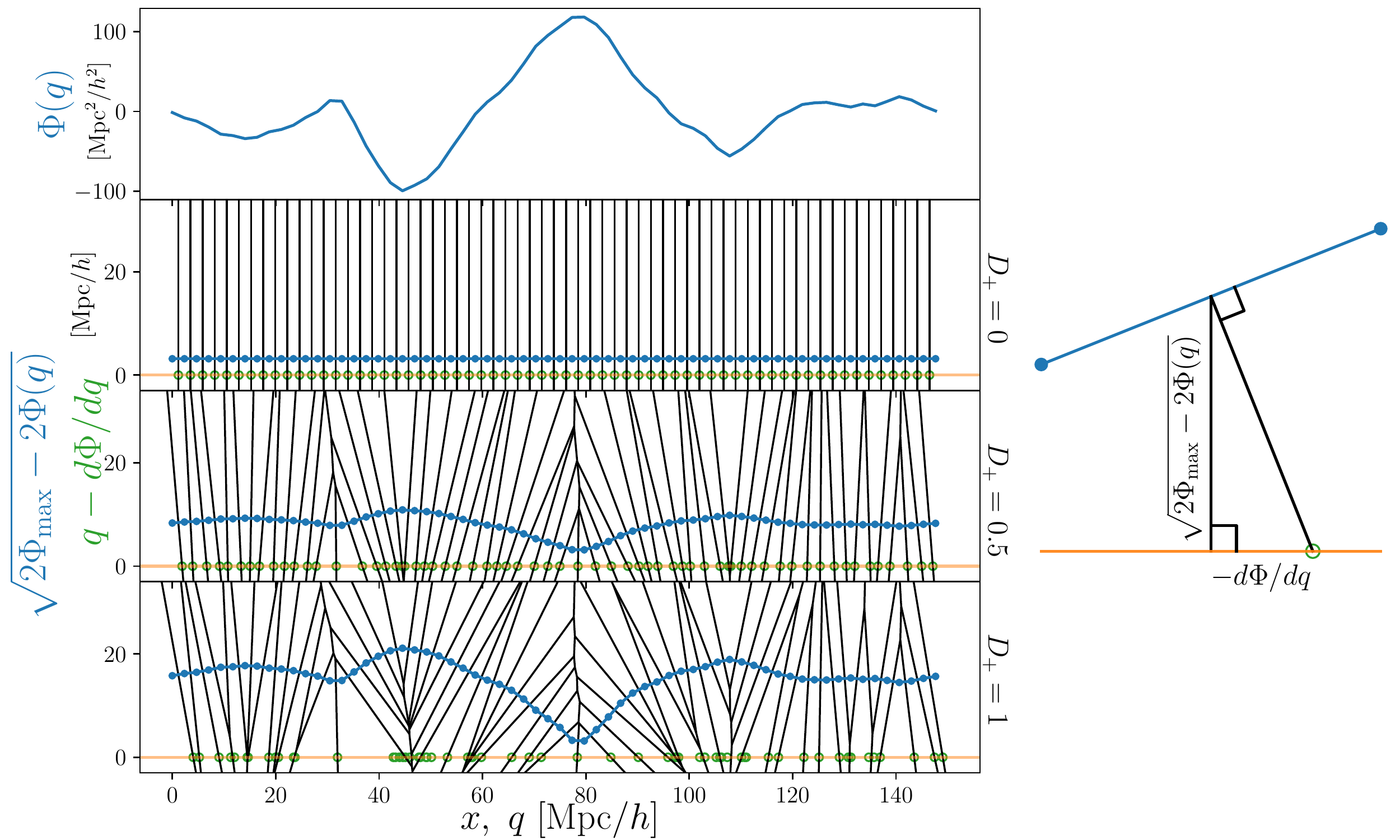}
  \end{center}  
  \caption{{\it Top-left}: An example 1D displacement potential. {\it Bottom-left panels}: Three snapshots of the evolution of the sectional-Voronoi tessellation. Generators at grid points begin at $\Phi(q)=0$, but as $\Phi(q)$ scales with the growth factor $D_+$, generators lift off the $x$-axis by a distance $\sqrt{2\Phi_{\max}-2\Phi(q)}$. The Eulerian cell of each grid point is the intersection of the $x$-axis with the full 2D Voronoi tessellation; `particles' are at these intersection points. In sufficiently deep potential wells, 2D Voronoi cells rise so high that the orange line no longer intersects them. Green circles show where the particles would be using the ZA, i.e.\ simply $x=q-\frac{d\Phi}{dq}$. In uncollapsed regions, the intersection points correspond to the circles, but note that in collapsed regions (e.g.\ at $x\approx 47$), green circles overcross. {\it Right}: Illustration of how the displacement field comes from this construction (see text).
  }
  \label{fig:sectional_voronoi}
\end{minipage}
\end{figure*}

The idea is easily visualized for a 1D universe. Fig.\ \ref{fig:sectional_voronoi} shows how linearly evolving a 1D potential in time changes its sectional-Voronoi tessellation, and 1D cosmic web. Each generator point is lifted to a height $h(q)=\sqrt{2\Phi_{\rm max} - 2\Phi(q)}$, for some $\Phi_{\max}\ge\ \max_{q} \Phi(q)$. (In Fig.\ \ref{fig:sectional_voronoi}, $\Phi_{\max}= \max_q \Phi(q) + 5$\hmpc, adding a constant for visual clarity.) If they did not collide, the black lines between generator points would all intersect the $x$ axis at $\Psi(q)=-d\Psi/dq$ (positions indicated with green circles). This is exactly the expected displacement in the ZA. To see this, consider the right panel of Fig.\ \ref{fig:sectional_voronoi}. A segment between two grid points of $h(q)$ is shown in blue. Its slope is simply the derivative of $h(q)$, $\frac{dh(q)}{dq} = -\frac{d\Psi}{dq}[2\Psi_{\rm max}-2\Phi(q)]^{-1/2}$. As it should, this expression agrees with the slope inferred using the right triangle in the figure.

In the left panels of Fig.\ \ref{fig:sectional_voronoi}, there is one Voronoi cell per generator point. The Eulerian cells in the 1D universe are the between intersections of the black lines with the orange $x$-axis. Where the potential is deepest, generators have floated up so high that their Voronoi cells no longer intersect with the orange line. This indicates collapse; these particles give their mass to the black line that does manage to intersect the orange line. We show the ZA position of these particles with green circles, using $x=q-\frac{d\Phi}{dq}$, evaluating the derivative with a finite difference between grid points. Note that in collapsed regions (e.g.\ at $x\approx 47$), green circles have some dispersion (having crossed each other), unlike the Voronoi edges in the collapsed regions that have adhered.

\begin{figure}
    \begin{center}
    \includegraphics[width=0.5\columnwidth]{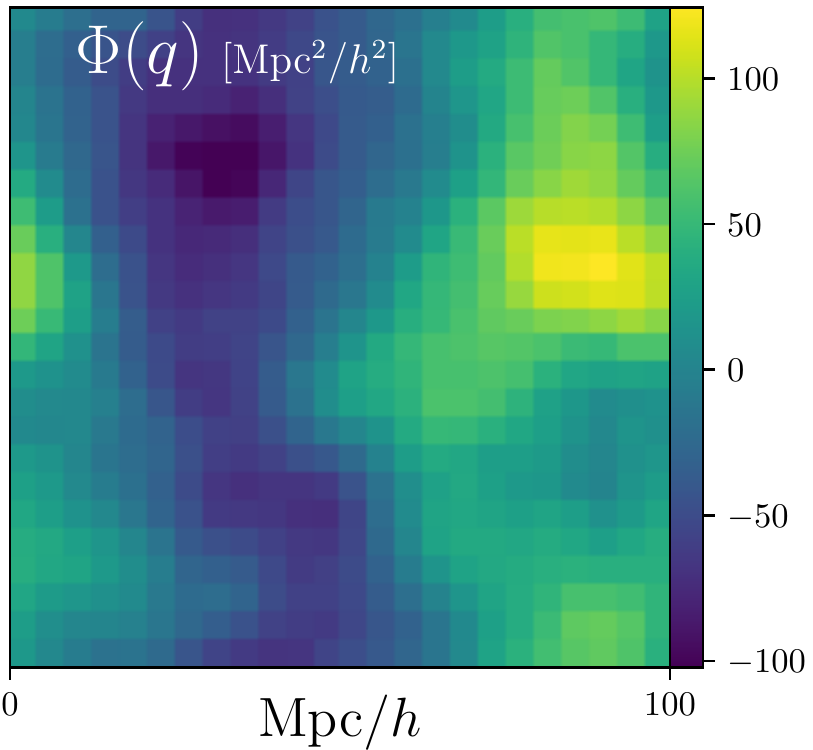}
  \end{center}  
  \caption{The displacement potential used to generate the following adhesion-model cosmic webs. It is $32^2$, extremely low-resolution for clarity.}
     \label{fig:displacement_potential}
\end{figure}
\begin{figure*}
 \begin{minipage}{175mm}
    \begin{center}
    \includegraphics[width=0.85\columnwidth]{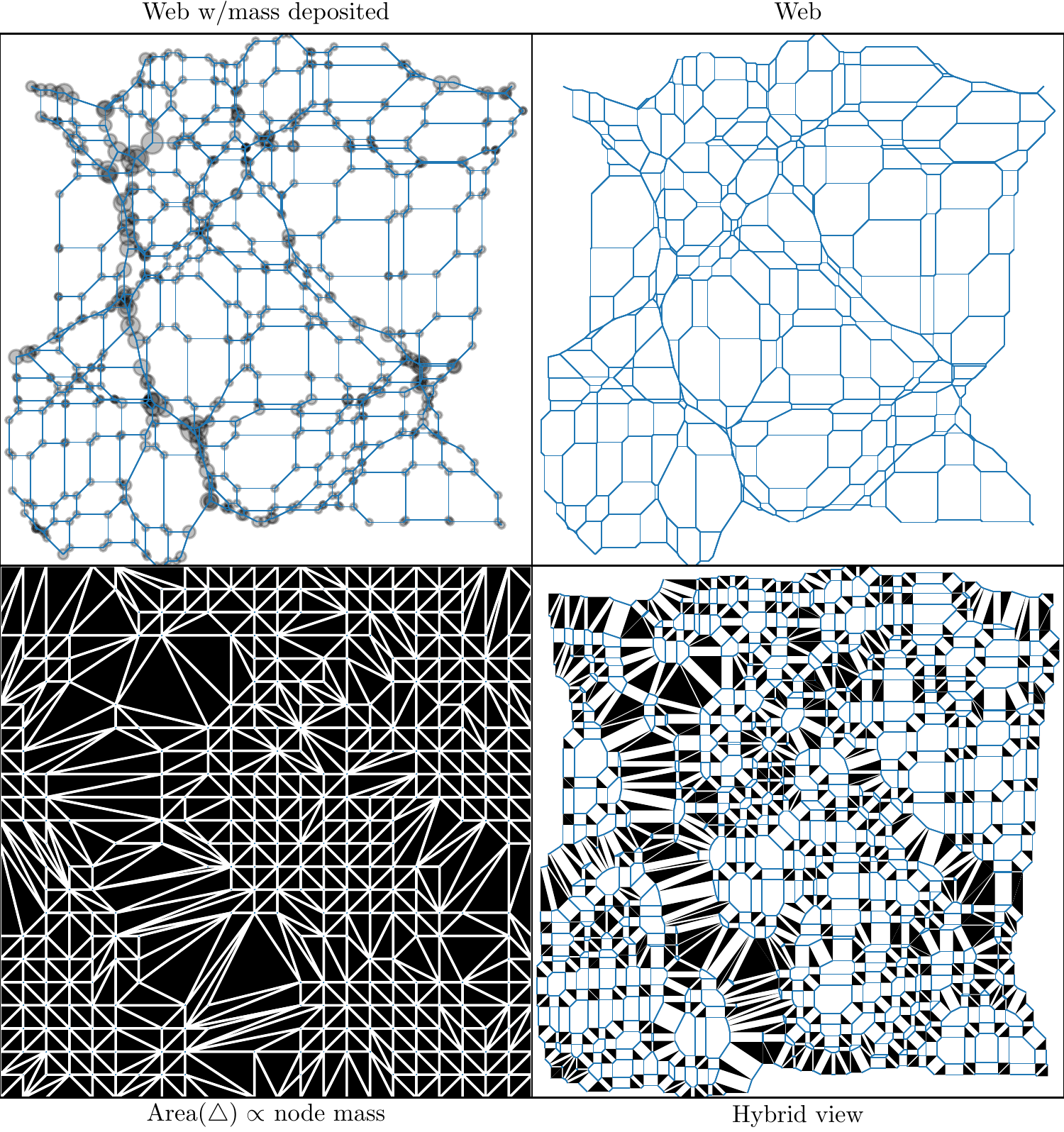}
    \end{center}
  \caption{{\it Upper right}: A cosmic web generated from the displacement potential in Fig.\ \ref{fig:displacement_potential}. Each blue polygon is a sectional-Voronoi cell (defined in Eq.\ \ref{eqn:secvoronoipot}), inhabiting Eulerian (final comoving position) space; the web collectively is a spiderweb. {\it Lower left}: The corresponding reciprocal dual tessellation, in Lagrangian (initial comoving position) space; each node of the Eulerian web is a triangle here. In architecture, the lengths of each white edges would be proportional to the tension in the corresponding spiderweb thread.  {\it Upper left}: the web at upper right, adding a translucent black circle at each node of area proportional to its mass (the area of its black triangle at lower left). {\it Lower right}: A Minkowski sum of the first two tessellations, every cell halved in linear size, i.e.\ $\alpha=\frac{1}{2}$ in Eq.\ (\ref{eqn:minksum}).}
    \label{fig:cosmicduals}
    \end{minipage}
\end{figure*}

\begin{figure}
    \begin{center}
    	\includegraphics[width=\columnwidth]{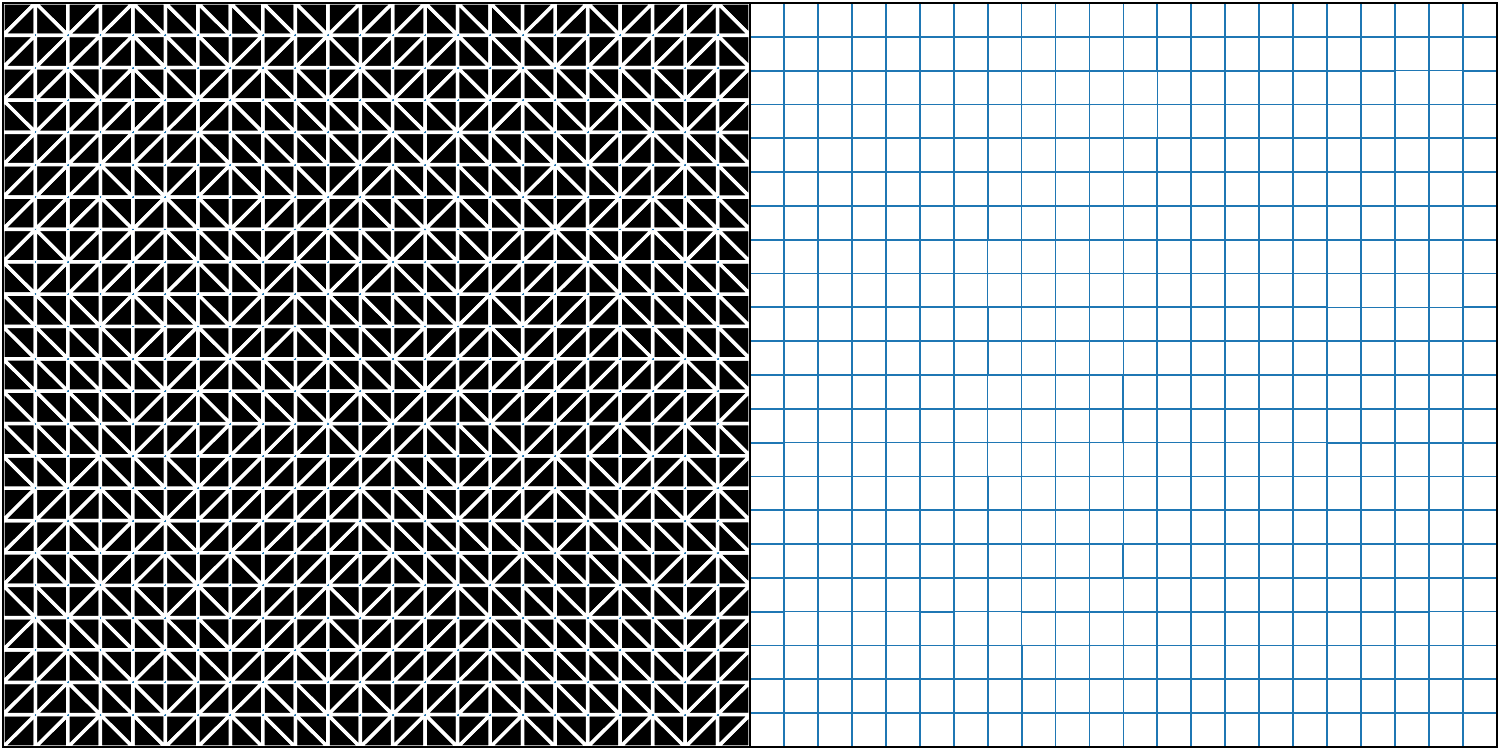}
    	\includegraphics[width=\columnwidth]{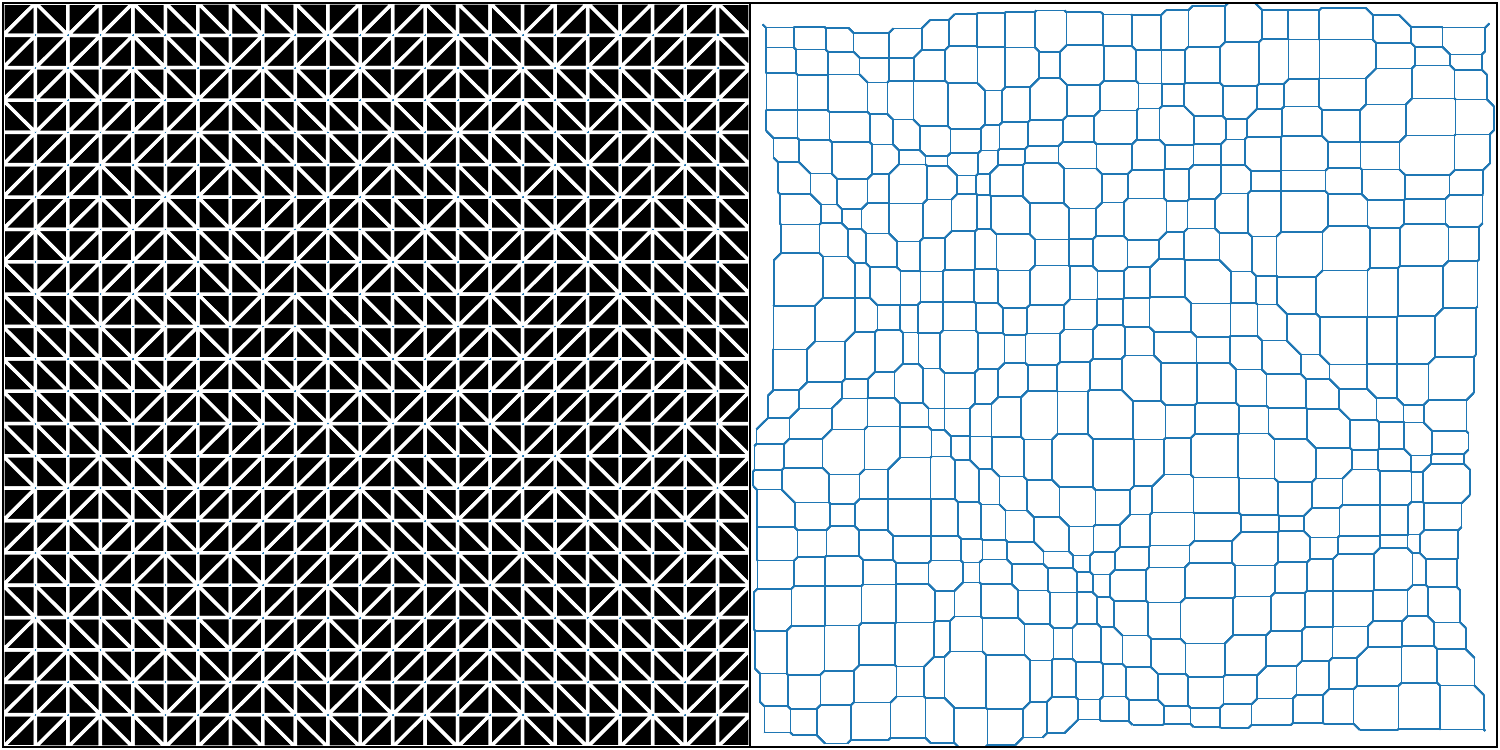}
    	\includegraphics[width=\columnwidth]{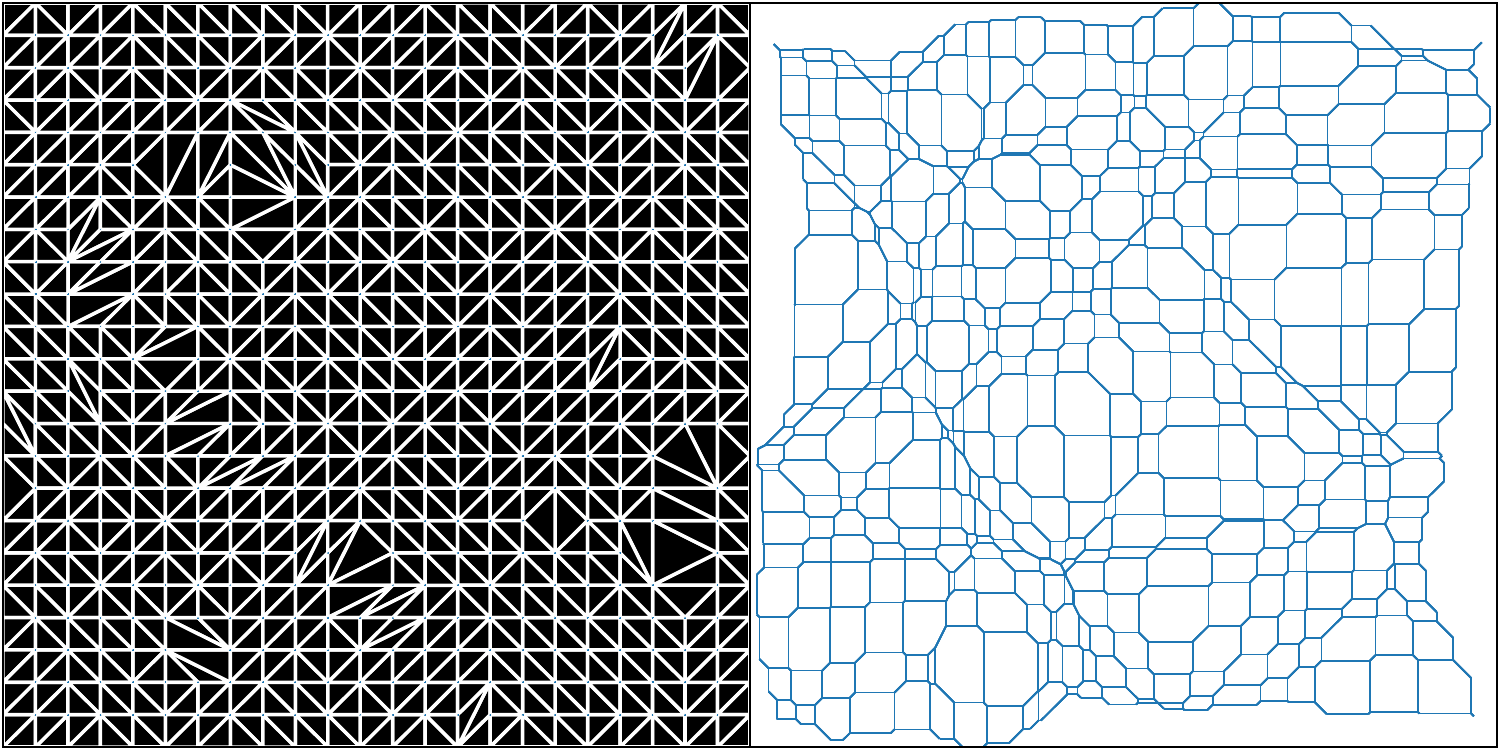}
    	\includegraphics[width=\columnwidth]{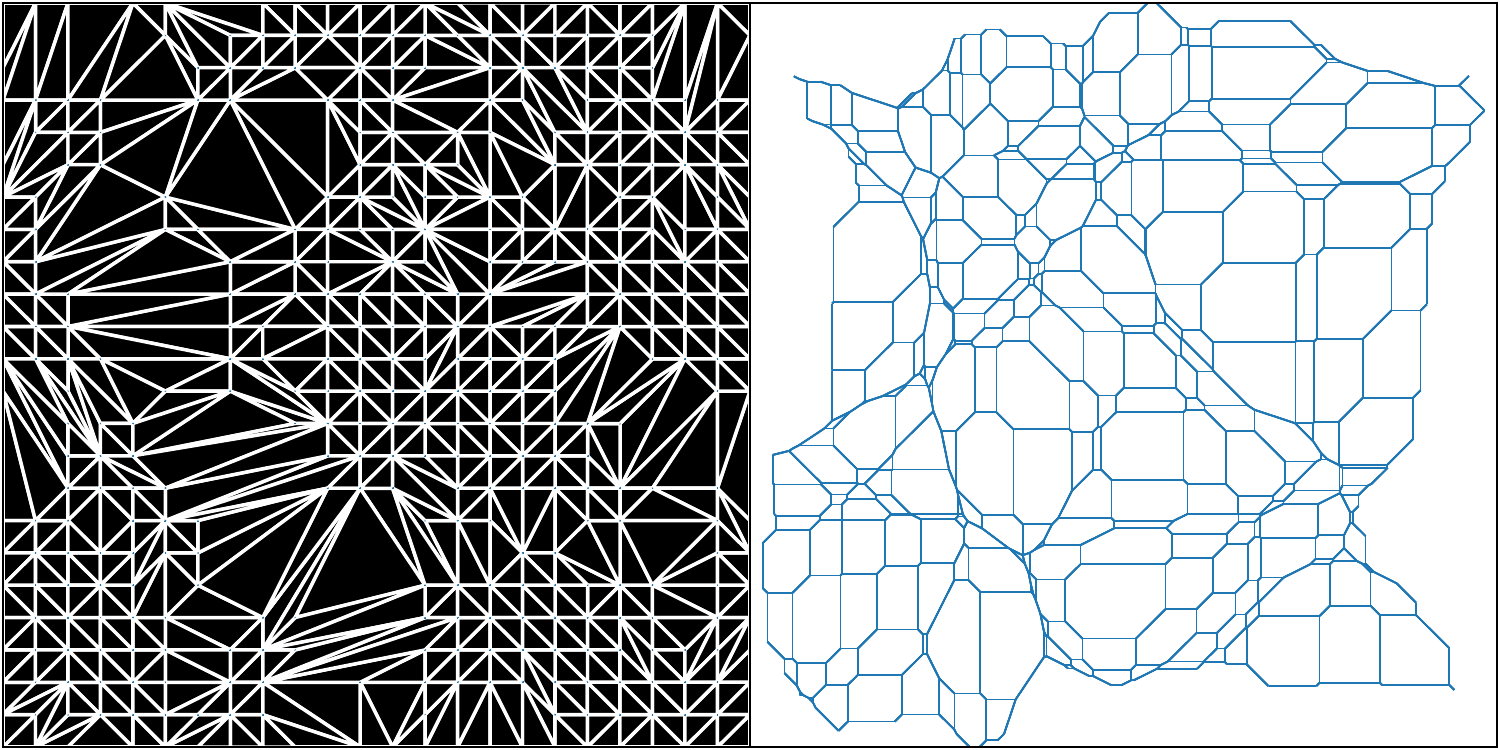}
  \end{center}  
  \caption{A time sequence of the adhesion-model cosmic web in Fig.\ \ref{fig:cosmicduals}, scaling the displacement potential in Fig.\ \ref{fig:displacement_potential} by $D_+=0,0.25,0.5$, and $1$, from top to bottom. {\it Left}: Lagrangian triangulation; each black patch collapses into a node of the web at right.}
  \label{fig:timesequence}
\end{figure}
The procedure to produce an adhesion-model realization of particles (in 1D, 2D or 3D) at scale factor $D_+$ is as follows:
\begin{enumerate}
\item{Generate a grid of initial conditions $\delta_0(\bq)$, i.e.\ a Gaussian random field consistent with an initial power spectrum, dependent on cosmological parameters.}
\item{From $\delta_0(\bq)$, obtain the displacement potential in Lagrangian coordinates, $\Phi(\bq)$. In the ZA, $\Phi(\bq)=D_+ \nabla^{-2}\delta_0$, the inverse Laplacian straightforwardly calculable with a fast Fourier transform (FFT).}
\item{Construct sectional-Voronoi and Delaunay tessellations from the grid. To do this, put a generator point at each grid point, setting the additive weight in the distance function for each generator point to $z^2(\bq) = -2\Phi(\bq)$. Each cell $V_{\bq}$ then satisfies
\begin{equation}
V_{\bq}=\left\{\bx\ {\rm s.t.}~ |\bx-\bq|^2 - 2\Phi(\bq) \le |\bx-\bp|^2 - 2\Phi(\bp),~\forall \bp\in \mathcal{L}\right\},
\label{eqn:secvoronoipot}
\end{equation}
where $\mathcal{L}$ is the set of Lagrangian grid points. (To make the tessellation truly a geometrical section of a higher-dimensional Voronoi tessellation, add $\Phi_{\rm max}$ to $-\Phi(\bq)$ as in the 1D example above, but this is not necessary computationally.) If the volume of $V_{\bq} = 0$, the Lagrangian patch at $\bq$ is part of a collapsed wall, filament, or node. If the volume of $V_{\bq}>0$, the patch at $\bq$ is uncollapsed. The mass of each particle in the web is given by the area of the corresponding triangle or tetrahedron of the dual tessellation (which is a reciprocal dual). We call this dual a `weighted Delaunay tessellation' (also known as a `regular triangulation'), a tessellation of the Lagrangian space of the initial conditions.}
\end{enumerate}

\subsection{2D example}
A $\Phi(\bq)$ field appears in Fig.\ \ref{fig:displacement_potential}, and the resultant cosmic spiderweb in Fig.\ \ref{fig:cosmicduals}. Each triangle in Lagrangian space, with mass given by its area (lower left) is a node of the spiderweb in Eulerian space (upper right). The nodes are shown with mass deposited at upper left, and a half-half Lagrangian/Eulerian mixture is at lower right, called a Minkowski sum.

A Minkowski sum of two sets of vectors $\bA$ and $\bB$ is $\bA+\bB \equiv \{\ba+\bb~|~\ba\in\bA, \bb\in\bB\}$. We alter this concept a bit, following \citett{McRobie2016}, for application to dual tessellations. We do not want to attach arbitrary vectors in each set to each other, but specify that there is a subset $\bB_i$ of the dual tessellation $\bB$ attachable to each vector $\ba_i\in\bA$. We also add a scaling $\alpha$ to interpolate between the original and dual tessellations. The vertices of our {\it Minkowski sum} satisfy
\begin{equation}
\alpha\bA+(1-\alpha)\bB\equiv \left\{\alpha\ba_i+(1-\alpha)\bb_j~|~\ba_i\in\bA, \bb_j\in\bB_i\right\}.
\label{eqn:minksum}
\end{equation}
The vectors in the sum are what we actually plot. Only spiderweb networks have Minkowski sums with parallel lines separating neighboring polygons of each tessellation like this.

Fig.\ \ref{fig:timesequence} shows the time evolution of this cosmic web, from uniformity at $D_+=0$, to the snapshot in Fig.\ \ref{fig:cosmicduals}. When triangles in the Lagrangian triangulation merge, the nodes stick together.

Some aspects of each tessellation may look curious. In the Lagrangian triangulation, why is there a square grid, with each square split by 45\degr, randomly-oriented diagonals? And why are the threads of the Eulerian web only horizontal, vertical, and 45\degr-diagonal in uncollapsed regions? These are both results of picking a square grid for the sectional-Voronoi generators.  The grid of Voronoi generators could just as easily form a triangular grid, or even an irregular, e.g.\ `glass' set of initial conditions. In that case, the threads outside collapsed regions would have random directions. Another possibly surprising thing is that opposite sides of the tessellation do not perfectly fit together if tiled, as expected from a periodic displacement potential. This is because it is missing boundary cells in such a repetition, cells between generators along the top and bottom, and the left and right.

So, in summary, what does it mean to say that the cosmic web in the adhesion model is a spiderweb, in 2D? A realization of the universe in the adhesion model, truncating the structure at a fixed resolution, consists of a set of particles of possibly different mass, at vertices of a sectional-Voronoi tessellation. If the edges of sectional-Voronoi cells are replaced with strands of string, this network can be strung up to be entirely in tension. The tension in each strand will be proportional to the length of the corresponding edge of the weighted Delaunay triangulation, which in cosmology is proportional to the mass per unit length along the filament. So, to construct this with the optimal amount of constant-strength material to be structurally sound, strand thicknesses would be proportional to their thicknesses in the Minkowski sum.

Note the tree-like appearance in the bottom-right panel of Fig.\ 5. In a tree (a structural spiderweb), the summed cross-sectional area of the trunk is roughly conserved after each branching. In our case, similarly, the total cross-sectional area (mass) of a large filament is conserved after branching. Referencing actual arachnid spiderwebs, gravity is like a haunted-house explorer, clearing strands aside, causing them to adhere and produce thicker strands.  Note that there are other networks in nature with approximate conservation of summed thickness across branches, such as biological circulatory networks, and traffic-weighted maps of traffic flow in and out of a hub \citepp{West2017}.

However, there is an aspect of this picture that is a bit at odds with a particular definition of the cosmic web as consisting of multistream regions \citepp[e.g.][]{FalckEtal2012,Neyrinck2012,RamachandraShandarin2015,RamachandraShandarin2017,ShandarinMedvedev2017}. This multistream picture of the cosmic web is particularly simple to relate to the adhesion model, but there are alternative definitions of the cosmic web that have their own advantages \citep{LibeskindEtal2017}. To be guaranteed of a cosmic web that is a spiderweb, uncollapsed as well as collapsed nodes of the sectional-Voronoi tessellation must be included. {\it Uncollapsed} nodes have not joined with other nodes; these are the smallest black triangles in Figs.\ \ref{fig:cosmicduals} and \ref{fig:timesequence}. {\it Collapsed} nodes are those that have joined with other nodes in the adhesion model; these would be expected to have experienced stream-crossing in full gravity. The multistream picture of the cosmic web is that set of only collapsed regions.

The distinction between the cosmic web of collapsed and uncollapsed nodes is relevant for the conceptual question of whether the uncollapsed/single-stream/void region {\it percolates} (i.e.\ is a single connected region) throughout the Universe, or whether multistream boundaries pinch it off into discrete voids. If uncollapsed nodes are included in the adhesion-model cosmic-web network, the set of `voids' is simply the set of all sectional-Voronoi cells. Each of these is its own exact convex polyhedron, by construction, and they do not percolate. More interesting, though, are the percolation properties of the region outlined by collapsed nodes; in $N$-body simulations, it seems that this region does percolate \citepp{FalckNeyrinck2015,RamachandraShandarin2017}.

A relevant question for further study is to what degree the cosmic web remains a spiderweb if uncollapsed nodes are excluded.  Imagine stringing up the blue Eulerian web in Fig.\ \ref{fig:cosmicduals}, to be in tension. To what degree could the web of collapsed nodes be preserved after clipping away edges in void regions? The shapes would generally change, e.g. some bends in filaments would straighten. But if the threads are carefully chosen, the change in shape could be small. Also, there is a freedom to change the initial arrangement of nodes away from a lattice, that be used to optimize the spiderwebness of the modeled cosmic web.

Note too that if the multistream region is not a single connected structure (i.e.\ if it does not percolate), then it cannot collectively be a spiderweb, although each connected patch could be. Thankfully (for cosmic-spiderweb advocates), \citett{RamachandraShandarin2017} find indeed that the multistream region generally nearly percolates if the mass resolution is reasonably high, at the current epoch. But this picture would be different at an early epoch, when the multistream region would indeed consist of different discrete isolated patches, that could individually but not collectively be a spiderweb. While the single-stream/void region appears to percolate at all epochs, the epoch when the multistream region percolates in the observable universe may define an interesting characteristic time for structure formation in the Universe.


\section{Origami tessellations}
Another physically meaningful way of representing the cosmic web is with origami. In fact, it was the origami result by \citett{LangBateman2011} that led us to the current paper. They give an algorithm for producing an {\it origami tessellation} from a spiderweb. Here we define an origami tessellation as an origami construction based on a polygonal tessellation that folds flat (in 2D, can be pressed into a book without further creasing). Also, we require that folding all creases produces a translation, but not a rotation or reflection, in each polygon of the original tessellation. The no-rotation property is usual but not universal in paper-origami tessellations. But it is crucial in the cosmological interpretation: single-stream regions, that correspond to these polygons, should hardly rotate with negligible large-scale vorticity.

By default, we mean `simple twist fold' tessellations, in which each polygon has a single pleat between them, and each polygon vertex has a polygonal node. A {\it pleat} is a parallel pair of creases; since they are parallel, neighboring polygons do not rotate relative to each other. Another relevant type of tessellation is a `flagstone' tessellation; \citett{Lang2015} also worked out an algorithm to produce one of these from a spiderweb. Polygons in a flagstone tessellation are designed to end up entirely top after creases are folded, requiring at least four creases between each polygon if flat-folded. Voronoi tessellations, reciprocal diagrams, and related concepts are rather widely used in origami design and mathematics \citepp{Tachi2012,DemaineEtal2015,Mitani2016}, e.g.\ playing a central role in the recent first `practical' algorithm for folding any polyhedron from a 2D sheet \citepp{DemaineTachi2017}.


\begin{figure}
	\begin{center}
	    \includegraphics[width=\columnwidth]{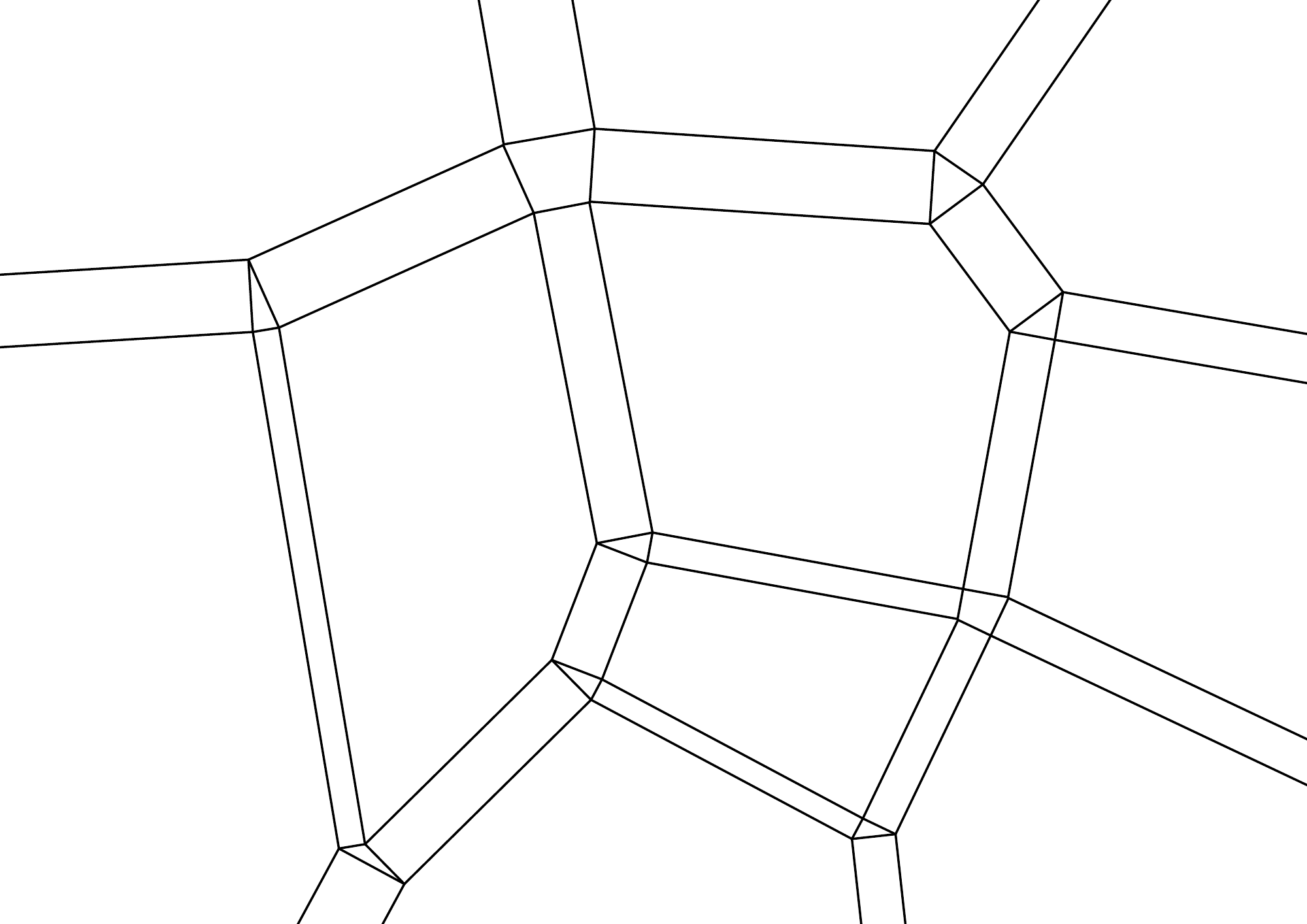}
	    	    \includegraphics[width=\columnwidth]{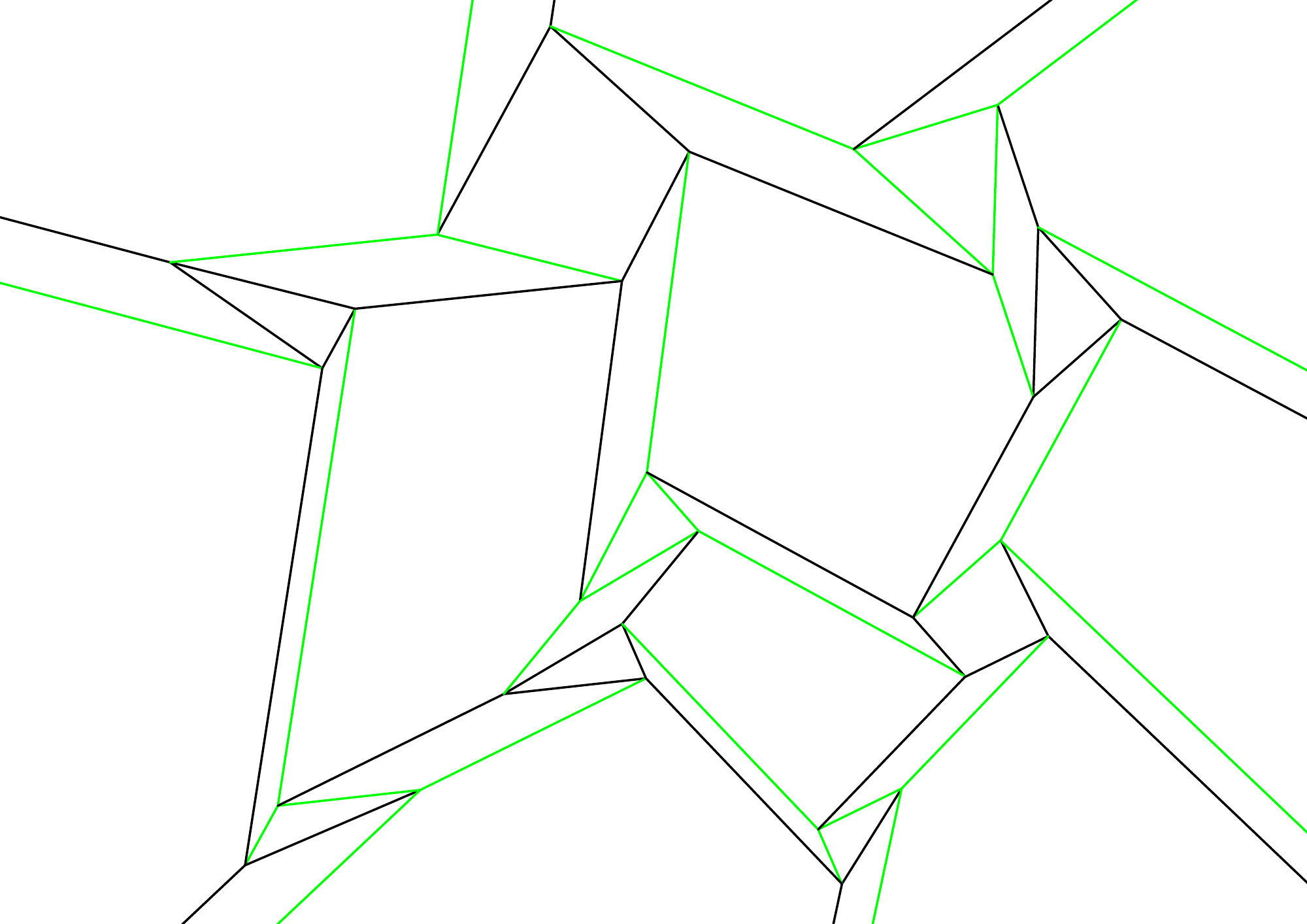}
	    \includegraphics[width=\columnwidth]{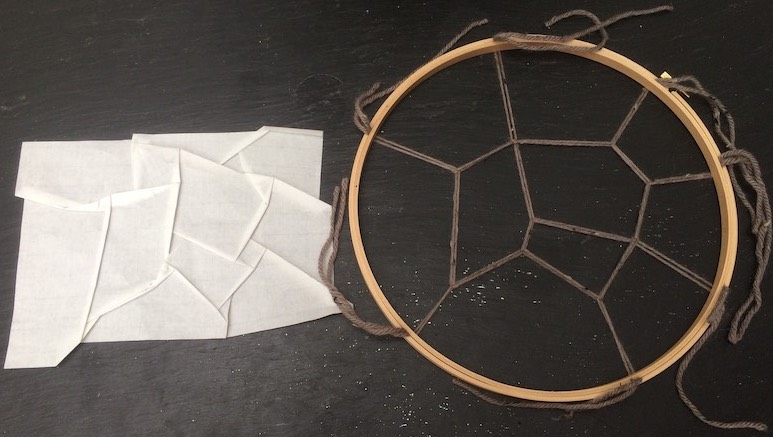}
    	\end{center}
  \caption{Origami and textile approximate representations of the Council of Giants \citepp{McCall2014}. {\it Top}: Eulerian-space Voronoi polygons, and dual, Lagrangian-space Delaunay polygons, combined in a Minkowski sum construction.  {\it Middle}: The same, with both tessellations rotated with respect to each other. With its smaller angles, it is actually foldable. Green lines are valley folds (looking like a V from the side when folding); black lines are mountain folds (looking like half of an M). {\it Bottom}: The middle panel folded from paper, alongside a nearly matching spiderweb construction built from yarn and an embroidery hoop. The length of a Delaunay edge at top is proportional to the tension in the corresponding strand of the spiderweb.}
  \label{fig:councilofgiants}
\end{figure}

Fig.\ \ref{fig:councilofgiants} shows an example of an origami tessellation, in a pattern designed to resemble the set of galaxies in the so-called Council of Giants \citepp{McCall2014}. These are `giant' galaxies (e.g.\ the Milky Way and Andromeda) within 6 Mpc of the Milky Way; all such galaxies happen to be nearly coplanar, along the Supergalactic Plane. Such a nearly 2D structure is particularly convenient to represent with 2D paper. We designed this by hand, from a Voronoi tessellation, with Robert Lang's {\it Tessellatica} Mathematica code.\footnote{\hrefurl{http://www.langorigami.com/article/tessellatica}}

The top panel shows both the web and its dual together, constructed as a Minkowski sum. The middle panel shows essentially the same, except Delaunay polygons from the top have has been rotated by 70\degr, and the Voronoi polygons shifted around accordingly, as prescribed by the `shrink-and-rotate' algorithm \citepp{Bateman2002}. In fact, the top panel was generated in the same way, but it is not foldable as actual origami without patches of paper passing through each other, because of the large, 90\degr\ angle between Delaunay and Voronoi polygons. The bottom panel shows the middle pattern folded from actual paper, alongside a textile spiderweb construction with the same structure.

Comparing this structure to the actual Council of Giants arrangement \citepp[][Fig.\ 3]{McCall2014}, it is only approximate; some galaxies are lumped together. All galaxies could have been included, but this would have required dividing up the voids, since each galaxy must have at least 3 filaments. Also, greater accuracy would have been possible by using different  weights in a sectional-Voronoi diagram; here, we used a simple 2D Voronoi diagram.

\subsection{Origami tessellations and the cosmic web}
Does an origami tessellation have a physical correspondence to the cosmic web? Yes, approximately. An origami tessellation describes the arrangement of dark matter after gravity has caused it to cluster, under the strict `origami approximation' \citepp{Neyrinck6OSME2015,NeyrinckZELD2016}. The strict part of this approximation is the requirement that the sheet does not stretch non-uniformly (i.e.\ stays constant-density, except for piling up when it is folded). In full gravity, this requirement is wildly violated, although in a way, it holds surprisingly well into multistream regions; \citett{VogelsbergerWhite2011} found that the median density on each stream is near the background density, even deep in the heart of a dark-matter halo (a collapsed dark-matter node that might be big enough to pull gas into it to form a galaxy), with up to 10$^{14}$ streams coinciding in some locations.

In the origami approximation, the density at a location after folding is the density of the sheet times the number of layers that overlap there. Imagine shining a light through the paper after folding, and interpret the darkness at each point as a density; the light freely shines through single-layer regions, but is blocked progressively more as the number of layers increases.

\begin{figure}
	\begin{center}
	    \includegraphics[width=0.72\columnwidth]{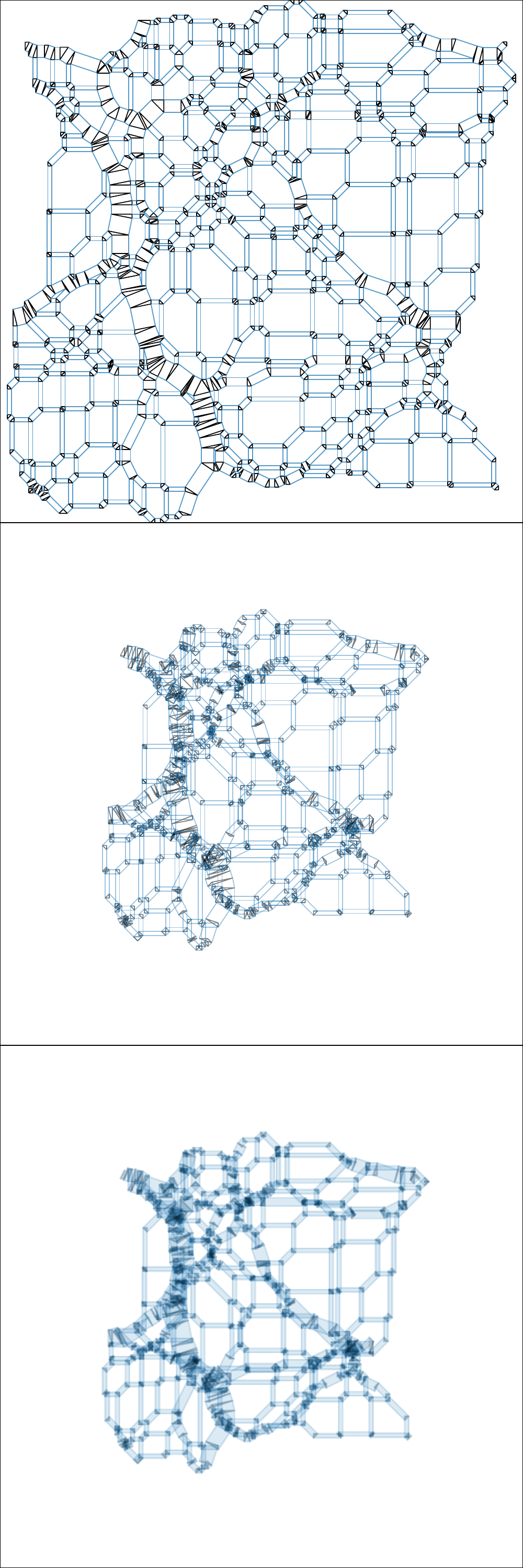}
    	\end{center}
  \caption{{\it Top}: An origami crease pattern, formed combining the edges of the sectional-Voronoi (blue) and Delaunay (black) tessellations of Fig.\ \ref{fig:cosmicduals} in a construction known as a Minkowski sum: every blue polygon is scaled by $\alpha=\frac{3}{4}$ and placed at a black node, and every black polygon is scaled by $(1-\alpha)$ and placed at a blue node. {\it Middle}: The locations of the creases after folding up the crease pattern, i.e. performing a reflection operation in each line segment. Note that the sheet has contracted, because patches (still the same size) now overlap. {\it Bottom}: The `density' in the folded-up middle panel, showing the number of streams (layers of the sheet) at each location, as though backlit. White polygons are single-stream, and the number of streams increases with the darkness of the color.}
  \label{fig:origamiweb}
\end{figure}

Since an adhesion-model cosmic web gives a spiderweb, and a spiderweb gives an origami tessellation, the adhesion model can give an origami tessellation, as in Fig.\ \ref{fig:origamiweb}. Does the folded-up origami tessellation in the bottom panel have a precise physical meaning for an adhesion-model cosmic web? 

Yes, with some caveats. Although the origami construction gives a good qualitative representation of the web, it does not necessarily give a good representation of the phase-space structure of the dark-matter sheet \citepp{ShandarinEtal2012,AbelEtal2012,FalckEtal2012,Neyrinck2012,HiddingEtal2014,FeldbruggeEtal2017}, the original inspiration for the origami construction. The {\it dark-matter sheet}, or Lagrangian submanifold, is a construction that keeps track of the way dark matter fills and moves in space. The dark-matter sheet is a 3D manifold folding without crossing itself or tearing in 6D position-velocity phase space. Its vertices, corresponding to particles in the usual conception of a cosmological $N$-body simulation, begin with tiny velocities and displacements away from a uniform grid. Then, starting from these tiny perturbations, gravity causes structures to fold up in 6D phase space.

Some aspects of this folding in a large-scale structure context are described by catastrophe theory; see \citepp{ArnoldEtal1982}, and recently \citepp{HiddingEtal2014,FeldbruggeEtal2017}. In fact, this link through catastrophe theory brings in interesting analogies to other physical systems. One example is the set of light caustics at play on the bottom of a swimming pool: an incoming sheet of light gets bent and folded into a pattern of light resembling a two-dimensional cosmic web \citepp{ShandarinZeldovich1984,ShandarinZeldovich1989,HiddingCaustics2014}. Often in nature and art, catastrophe theory deals with smooth curves, as emphasized by \citepp{McRobie2017seduction}. Compared to these, the origami and spiderweb constructions discussed here are angular, composed entirely of straight lines. The reason for this difference is that we are considering systems such as the cosmic web at a fixed resolution, at which the structure is angular. The physical, curved edges of structures would reveal themselves if the resolution is increased.

In the Minkowski sum construction, folding occurs even at uncollapsed nodes, which correspond to single resolution elements. But no folding of the phase-space sheet is expected to occur there. At collapsed nodes, though, the representation holds reasonably: the sheet does fold qualitatively like it would in phase space. The larger the node, the more massive it is physically.

What about the correspondence between origami pleats/filaments and more usual conceptions of cosmic filaments? This is an important question for us, since the calculation that originally inspired the term `cosmic web' \citepp{BondEtal1996} showed that typically, from the initial conditions onward, there is an overdense column of matter (a filament) between nearby collapsed peaks. Origami pleats/filaments do indeed correspond rather well to cosmic filaments. Sensibly, the thickness of each origami pleat/filament is proportional to the length of the edge it corresponds to in the Lagrangian triangulation; this indicates its mass per unit length if interpreted cosmologically.

But, reducing the physicality of the origami construction, pleats exist in uncollapsed void regions as well (but are only one resolution element wide). These fold up in the origami construction, but do not correspond to multistream regions. And for a wide, physical filament in the origami construction, whenever it bends, it must do so with a node, joined to at least one additional pleat. If that additional pleat is just one resolution element wide, it does not correspond to a physical collapsed filament, and the node producing the bend in the larger filament does not correspond to a physical node, i.e.\ a structure that has collapsed along more than one axis.

Another way the origami folded model could be relevant physically is in estimating the density field from an adhesion-model realization. But the question of how most accurately to assign density to an Eulerian grid given an adhesion-model realization is beyond our current scope.

\section{Three dimensions}
What is a 3D spiderweb? The typical picture of a biological spiderweb is nearly planar, but many spiders spin fully 3D webs (such as the black widow; see Tom\'{a}s Saraceno's {\it 14 Billions}). The concept of cosmic webs in the adhesion model, and the structural-engineering spiderweb, carry to 3D as well. Indeed, the field of fully 3D graphic statics that employ concepts such as the reciprocal diagram has experienced a resurgence of interest, and is currently an active area of research.

Most of the spiderweb concepts we have discussed generalize straightforwardly to 3D. A 3D spiderweb is a network of members joining nodes that can be strung up to be entirely in tension. \citett{Rankine1876} introduced the concept of a reciprocal dual in 3D. The following will mirror the 2D definition. Consider a 3D spatial graph of positioned nodes, and edges between them. A {\it dual} to this graph is a tessellation of space into closed polyhedral cells, one per original node, such that neighboring nodes of the original are separated by faces in the dual. A dual is {\it reciprocal} if each original edge is perpendicular to its corresponding face in the dual. The original graph is a {\it spiderweb} if a reciprocal dual exists such that its cells fit together at nodes with no gaps or overlaps between them.

In structural engineering, the form diagram of a 3D truss is the map of members and nodes in space, and the force diagram is a collection of fitted-together {\it force polyhedra}, one polyhedron per node. If a node's force polyhedron is closed, it is in force equilibrium, and the forces on the each member meeting at the node is proportional to the area of the corresponding face of the force polyhedron.

What about the 3D cosmic web in the adhesion model? As in 2D, Eulerian space gets tessellated into polyhedral cells of a sectional-Voronoi diagram according to Eq.\ \ref{eqn:secvoronoi}, the additive weight for each cell given by the displacement potential at that point. And the volume of a tetrahedral cell in the weighted Delaunay tessellation of Lagrangian space gives the mass which contracts into the corresponding node.

The `cosmic web', i.e. the spatial graph of sectional-Voronoi edges that inhabit Eulerian space, is a spiderweb in 3D as well as in 2D, since a non-overlapping reciprocal dual tessellation exists: the faces of the Lagrangian weighted Delaunay tessellation are perpendicular to the corresponding Eulerian sectional-Voronoi (cosmic web) edges.

However, this definition of a `cosmic web' does not entirely conform with some common conceptions of the cosmic web. As in the 2D case, the cosmic spiderweb includes uncollapsed nodes in single-stream regions, usually classified to be in voids \citepp[e.g.][]{FalckEtal2012,RamachandraShandarin2015}. Also, while edges of the sectional-Voronoi tessellation can be identified with filaments, we have not mentioned a similar concept for the walls. Perhaps if the faces of the sectional-Voronoi faces were filled in panels, these panels instead of the beams could provide an alternative structural-engineering description of the cosmic web (a cosmic foam, instead of web).

There is another subtlety to this correspondence in 3D. While in 2D, the definitions of spiderwebs and cosmic webs are exactly the same, both arising from a sectional-Voronoi description, so far the correspondence in 3D is only rigorous in one direction: cosmic webs are spiderwebs, but it seems that not every structural-engineering spiderweb can be constructed from a sectional-Voronoi diagram. \citett{McRobie2017} explains (see e.g.\ Fig. 8) that there exist three-dimensional structural-engineering spiderwebs that are not sectional-Voronoi diagrams (and therefore cosmic webs). This is because in structural engineering, it is only the force on a structural member that matters (i.e. the area of the face in the force diagram), not its shape, but the shapes must also match when connecting nodes of a sectional-Voronoi diagram. However, this subtlety does not impact our main conclusion, that the cosmic web is a spiderweb.

\subsection{Origami tessellations in 3D}
As in 2D, 3D spiderwebs lead to `origami tessellations,' but not of paper, but of a non-stretchy 3D manifold, folding (being reflected) along planes in higher dimensions; when projected back to 3D, several layers of the manifold can overlap. The 3D origami crease patterns are again related to Minkowski sums; \citett{McRobie2016} gives several structural-engineering examples. Nodes are typically tetrahedra. A Toblerone-like, triangular-prism filament connects the faces of each neighboring pair of nodes, the filament's cross-section given by the triangle separating these tetrahedra in the Lagrangian Delaunay tessellation/force diagram. In structural engineering, the length of the filament gives the length of the structural member, and the force on it is proportional to the cross-sectional area. Between filaments are gaps (`walls') consisting of parallel identical polygons that are matching faces of neighboring cells of the sectional-Voronoi tessellation.

In cosmology, folding up a dark-matter sheet constructed in this way can give a way to estimate a density field made from a dark-matter sheet under the `origami approximation' \citepp{Neyrinck6OSME2015,NeyrinckZELD2016}. In cosmology, each surface represents a caustic. Each node would be a `3D twist fold' in origami terms, or `tetrahdedral collapse' \citepp{Neyrinck2016tetcol} in astrophysical terms. In the simplest, irrotational tetrahedral collapse, filaments extrude perpendicularly from faces of tetrahedral nodes. When collapse happens, nodes, filaments, and walls all collapse together. The four faces of each node invert through the node's center, the three faces of each filament invert through its central axis, and the two parallel faces of each wall pass through each other. Generally, rotation can happen; as a node collapses and inverts, it can undergo a 3D rotation as well. This causes its filaments to rotate, as well; this can correlated the rotation of nearby filaments.

However, as above in 2D, there are many cosmic-web nodes (haloes, in the origami approximation) that do not represent collapsed objects; they are just tracers of the structure in void, wall, or filament regions.  In the adhesion model, a cosmic-web classification (into voids, walls, filaments, and haloes) can be done according to the shapes of Lagrangian-space tetrahedra that collapse to form nodes. Haloes collapse from nearly equilateral tetrahedra (all sides many resolution elements long); filament segments collapse from slab-like tetrahedra (with only one side a single resolution element long); wall segments collapse from rod-like tetrahedra (with 2 or 3 sides a single resolution element long); and void patches are uncollapsed, all tetrahedron sides only a single resolution element long. Tetrahedron shapes indicate the directions along which mass elements have merged/stream-crossed. This is consistent with the \origami\ classification (\cite{FalckEtal2012}; \cite{NeyrinckEtal2015}; first introduced in \citealp{KnebeEtal2011}), of voids, walls, filaments and haloes according to the number of orthogonal axes along which streams have crossed.

\subsection{A 3D cosmic web in compression}
\begin{figure}
	\begin{center}
	    \includegraphics[width=\columnwidth]{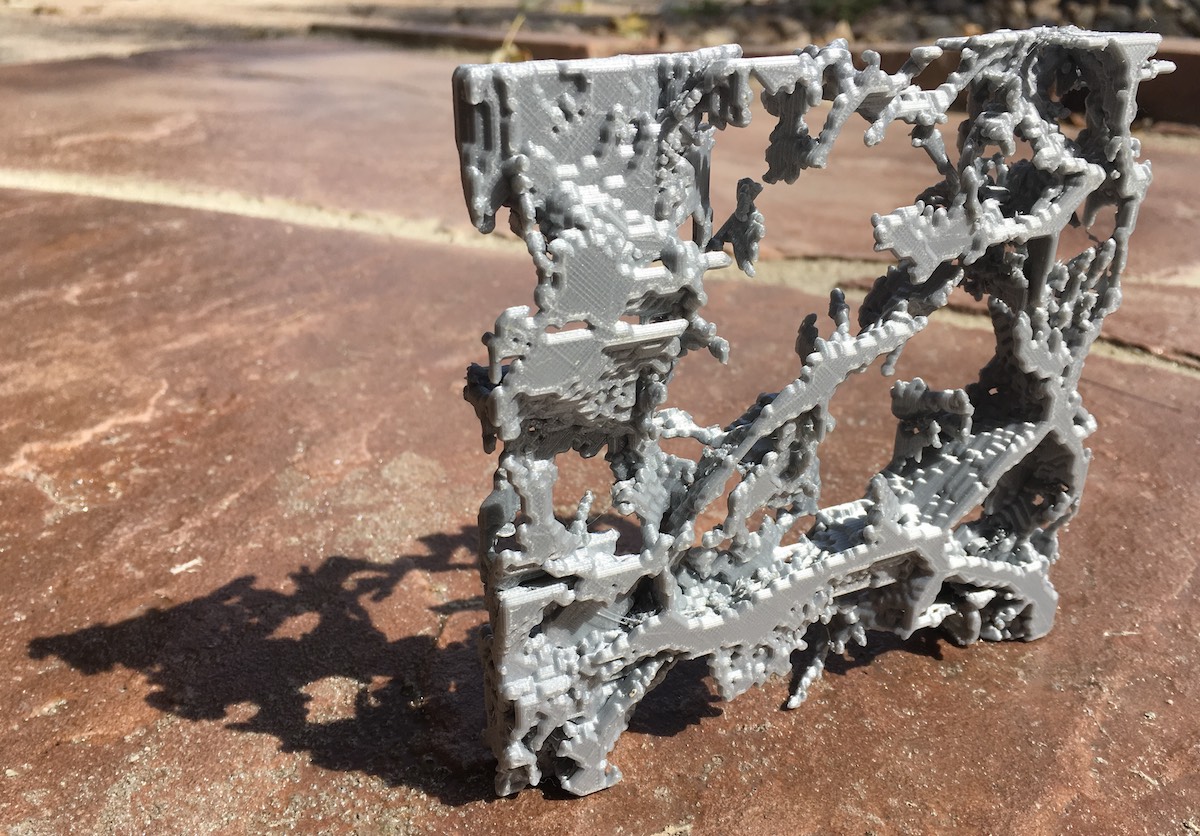}
    	\end{center}
  \caption{3D print made of plastic from a slice 50\hmpc\ on a side of an $N$-body simulation assuming a warm dark matter cosmology. This is a spiderweb mostly in compression (`mostly' because some of its segments are likely in tension). It is obviously strong enough to support its own weight; we did not test its breaking point, though.}
  \label{fig:threedprint}
\end{figure}

The everyday concept of a `spiderweb' refers to nodes and members in tension, but in structural engineering, it can just as well refer to members entirely in compression. Fig.\ \ref{fig:threedprint} shows a 3D-printed, tactile realization of a cosmic web. Its spiderweb nature simply means that it has structural integrity, and can support nonzero weight.

This model came from a 512$^3$-particle cosmological simulation with box size $100$\hmpc, assuming a warm-dark-matter cosmology. The initial conditions were smoothed \citepp[with parameter $\alpha=0.1$\hmpc; for details, see][]{YangEtal2015}, removing substantial small-scale structure, and simplifying the design. That is, the smallest dwarf galaxies observed may not form in this simulation. Note, though, that distance from the Milky Way to Andromeda would only be a couple of pixels, so differences within the Local Group would hardly show up. The main effect of the initial-conditions smoothing was to smoothen walls and filaments.

In detail, from the simulation, we pixelized a $50\times 50\times 6.25$ (\hmpcnosp)$^3$ volume with a $128\times 128\times 16$ grid (the mean number of particles per cell was 8). We filled in only multistream voxels, i.e.\ voxels containing collapsed (wall, filament or halo) particles as classified by \origami. We included only the largest connected set of such voxels within the slice. Of the 262144 voxels, 215127 had no collapsed particles and therefore were not in the structure, 46056 had collapsed particles and formed the structure, and 894 were collapsed but excluded because they were not connected within the slice (many of these would be connected through regions of the simulations outside the slice, though).

To produce the 3D print file, we follow \citett{DiemerFacio2017}, who make a 3D print similar in spirit to this, but higher resolution, and defined by a contour of a Gaussian-smoothed log-density field, not by multistreaming. We use the {\tt coutour3d} function in the {\scshape Mayavi} visualization package to produce a wavefront file from the contiguous set of printed voxels. We then import this into the freely available {\scshape Blender} package, and use its 3D printing toolkit plugin to remove possibly problematic artifacts for a 3D printer, keeping the appearance of the model the same. We then export an .stl file\footnote{\url{http://skysrv.pha.jhu.edu/~neyrinck/cosmicweb128.stl}} and print it.

This construction is likely a spiderweb, since it stands up under its own weight, but note that some of the members are likely in tension rather than compression, e.g.\ filaments that hang down and are not connected across voids. Some members in tension does not necessarily mean that an all-compression state cannot exist, though. We note again the `collapsed' vs 'uncollapsed' caveat: this would only be rigorously guaranteed to be a spiderweb if all void nodes were included (i.e., in the context of a simulation, all particles). In this case, this would fill in all but apparently random cells, giving an uninteresting design. But if the mass resolution (particle sampling) were decreased quite a bit, a 3D print including uncollapsed nodes could still be interesting (looking e.g.\ like a 3D version of the web in Fig.\ \ref{fig:cosmicduals}).

\section{Uses of the cosmic spiderweb picture}
\label{sec:applications}
If the cosmic spiderweb could be unambiguously identified observationally, analyzing it could possibly constrain cosmological models, in the ways we explain below.

But first we should clarify what an observation of `the cosmic spiderweb' means. Suppose the full dark-matter density field could be observed to some precision. The associated cosmic spiderweb would be a network of nodes and edges whose density field is consistent with these observations. This network would likely have nodes (essentially particles, in an $N$-body sense) in voids, to reproduce observed bends and kinks in walls and filaments. This procedure is essentially an inference of the initial conditions, for which many algorithms exist already \citepp[e.g.][]{KitauraEnsslin2008,KitauraEtal2012,Kitaura2013,HessEtal2013,JascheWandelt2013,LeclercqEtal2015,WangEtal2017,ShiEtal2017}. The sectional-Voronoi method for estimating the final structure is likely competitive with other fast approximations. But one unique aspect of the adhesion model is that its set of generating points can be placed on an irregular grid, and even with irregular initial generator masses. Arbitrary generator positioning would allow a better fit, but at the expense of vastly expanding the parameter space (a 3D position plus 1 velocity potential, per particle). The positions need not be entirely free, though; for instance, rather unconstrained regions such as voids could be traced with a minimal number of generator points.

An example would be to test how well the galaxies in the Council of Giants (a hand-designed model appearing in Fig.\ \ref{fig:councilofgiants}) can be reproduced with nodes of a sectional-Voronoi tessellation; galaxy masses and spins could even be added to the constraints. than in the hand-designed model in Fig.\ \ref{fig:councilofgiants}. A few issues made this more difficult than in a scientific test: we optimized for an appearance resembling typical conceptions of the cosmic web (notably, without nodes in voids, and minimizing the number of filaments). Also, the design algorithm uses pure Voronoi instead of sectional-Voronoi tessellations; a better fit would have been possible if displacement potentials as well as Voronoi generators could be tweaked.

But also, further constraints could make the test even better. Estimates of dark-matter halo masses (node areas or volumes in Lagrangian space), velocities, spins (as addressed in the next paragraph), and observations (or upper limits) on filaments between galaxies would all add useful information. As we plan to discuss in a future paper, the adhesion model could provide an elegant formalism for estimating spins in collapsed regions. Each collapsed node in the adhesion model collapses to a point from a patch of Lagrangian space, with some initial velocity field. Each patch would generally have nonzero angular momentum, even averaging from a potential velocity field. It is an open question whether and how this initial velocity field would continue to grow inside a collapsed structure, but perhaps the directions of galaxy spins would be accurately enough predicted in the adhesion model to provide useful constraints.

Here is a list of possible applications and an example of analyzing an observed cosmic web with these ideas. Most of these are simply aspects of the adhesion model that do not crucially relate to spiderweb ideas, but the last two items describe and give an example of a new geometric test.
\subsection{Experimentally testing structural integrity of tactile models}
The spiderwebness of an observed patch of the Universe could be tested by physically building a model of it and mechanically testing its structural properties, e.g., finding the weight it can bear before breaking. However, such a structural-integrity test is not quite a test for spiderwebness, since some members can be in tension if the object is put in compression. Indeed, the class of 3D-printed objects that can bear at least their own weight is of course quite broad, especially if the material is strong in tension as well as in compression. On the theoretical side as well, we must admit that the class of spiderwebs with as many connections between nodes as occur in the adhesion model is quite broad.

Still, building tactile models can often be surprisingly useful scientifically, for building intuition and flagging problems in the data. It is especially crucial for the visually impaired. And quantifying the structural integrity of the cosmic web could possibly provide scientifically useful constraints, but first, ambiguities would have to be cleared up. It is hard to imagine final constraints being derived from anything but entirely deterministic computer algorithms, but 3D-print-shattering experiments could be useful for intermediate results, and would obviously be delightful for educational and public-outreach purposes.

\subsection{Displacement fields without assuming periodicity}
 Typically, initial particle displacements in cosmological simulations are generated from a density field using an FFT, obtaining the displacement potential with e.g.\ the ZA. There are other, more accurate methods that work from a displacement-divergence (easily convertible to a displacement potential), that similarly use FFTs \citepp{Neyrinck2013, KitauraHess2013,Neyrinck2016}. It would be interesting to combine one of these methods with the adhesion model. These models implement a spherical-collapse prescription to prevent overcrossing, as the adhesion model does, but additionally predict void densities accurately.

The Voronoi method obtains particle displacements from the displacement potential without an FFT, which could be useful for investigations of flows on the largest observable scales. It also can naturally generate mutliresolution particle realizations, by varying the volumes occupied by particles in Lagrangian space. However, note that existing methods to generate a displacement potential (from e.g.\ a Gaussian random field giving the density) do use a periodic FFT.

\subsection{Identifying rotational or multistream displacements}
The adhesion model assumes a potential displacement field, i.e.\ $\bnabla_{\bq}\times\bPsi(\bq)={\bm0}$.  Collapsed regions in full gravity can carry quite large $|\bnabla_{\bq}\times\bPsi(\bq) |$. It is likely even nonzero but small \citepp{Chan2014,WangEtal2014} outside of collapsed regions, as in third-order Lagrangian perturbation theory \citepp{Buchert1994}. The degree of agreement with a spiderweb measures the magnitude of rotational vs irrotational motions, which might be a probe of the growth factor, and perhaps for modified gravity, some theories of which are known to affect the cosmic web \citepp[e.g.][]{FalckEtal2014,LlinaresMota2014,FalckEtal2015}.

Difficulty reproducing a high-precision galaxy arrangement with a sectional-Voronoi tessellation could be a signal of substantial rotational displacements. One cause could be stream crossing on few-Megaparsec scales, which in simulations seems to be a predictor for halting star formation via a `cosmic-web detachment' mechanism. Primordial filaments are thought to feed cold gas into galaxies, providing a fresh gas supply for star-formation; when these are detached, star formation is suppressed \citepp{AragonCalvoEtal2016}. Or, it could be a signal of unexpectedly high vorticity, or vector modes, in the initial conditions. Such a scenario is unlikely physically, since cosmic expansion is thought to dampen these away, but have received some consideration \citep[e.g.][]{Jones1971,ShandarinZeldovich1984} and unexpectedly high primordial vorticity is worth testing for.

Alternative methods to estimate the displacement curl exist, as well. If it were possible to estimate the full displacement field directly, one could measure its curl and divergence, immediately constraining these components. In principle, this could be done by fitting $N$-body simulations to observations. Note, though, that many shortcuts used for this task become unavailable if the displacement field is allowed to have a curl.

\subsection{Identifying anisotropy on large scales}
Spiderwebs, i.e.\ (sectional) Voronoi tessellations, are sensitive to anisotropies in a field. Voronoi tessellations cannot straightforwardly be used in a space where there is no global metric, e.g.\ position-velocity phase space in an $N$-body simulation \citepp{AscasibarBinney2005}. But this sensitivity can be exploited, as well. \citett{EvansJones1987} detect shear in a network of ice cracks, by fitting a Voronoi pattern to them. A Voronoi pattern fits an isotropic pattern well, but requires a scaling of the metric in one direction for a good fit to a sheared pattern.

In large-scale structure, applying these ideas would give a test similar in spirit to the Alcock-Paczynski (AP) test. Originally, \citett{AlcockPaczynski1979} proposed the test as a probe of the cosmological constant: if there existed `standard spheres' in the large-scale structure, their ellipticities could be used as a test of the expansion history assumed to go from redshift space to real space. Unfortunately, such `standard spheres' do not exist, but the AP test can be applied to contours of the redshift-space correlation function \citepp[e.g.][]{LiEtal2016}. Getting closer to the original spirit of the AP test, it can be applied to average redshift-space void profiles \citepp{Ryden1995,LavauxWandelt2012,HamausEtal2016,MaoEtal2017}. Redshift-space distortions spoil an idealized AP measurement, but simultaneously mix in a sensitivity to the growth rate of fluctuations.

In principle, departures from anisotropy in a cosmic spiderweb could be detected using each of its parts, without averaging structures. One way of doing this would directly use the key perpendicularity property (between the original and dual tessellation) of spiderwebs. If void `centers' (likely, density minima, i.e. generating points) could be observed, edges joining void centers should be perpendicular to walls and filaments between the voids. One could easily define a statistic quantifying this perpendicularity at each wall or filament. It could either be used in a position-dependent manner, or summed up over the survey for a global statistic quantifying the departure from a spiderweb. Notably, the perpendicularity test {\it carries no explicit cosmic variance}. This is in contrast to cosmological tests using correlation functions, power spectra, or even voids, in which even with perfect sampling of the density field, cosmic variance is present as fluctuations away from the cosmic mean. However, there will be noise in practice, causing constraints inferred from the perpendicularity test to get better as the volume is enlarged; the noise would likely behave in effect like cosmic variance.

This perpendicularity test would be highly sensitive to redshift-space distortions, just as in the AP test; redshift-space distortions substantially change the directions that filaments and walls have compared to real space. However, again, with sufficient modeling, this issue could be exploited instead of seen as an obstacle: the angles between void minima and walls between them could be used as position-dependent probe of redshift-space distortions. Still, there will likely always be various observational effects that add ambiguity in inferring void `centers,' as well as the positions and characteristics of walls, filaments, and haloes.

Note also that a section through a section is still a section; i.e. if the 3D cosmic web is a sectional-Voronoi tessellation, so will be a 2D slice through it. This makes these ideas applicable to large-scale structure surveys that are effectively 2D.

An alternative way of using the cosmic spiderweb to look for scalings of one spatial coordinate with respect to another is by analyzing the displacement potential inferred from an adhesion-model initial-conditions reconstruction.  Without scaling the metric according to the shear, a pure Voronoi tessellation cannot fit a sheared Voronoi tessellation \citepp{EvansJones1987}. Unfortunately for detecting shear in a sectional-Voronoi tessellation, though, a sectional-Voronoi tessellation can perfectly fit a sheared pure-Voronoi tessellation; a uniform shear can be produced by a large-scale gradient in the generators' weights.\footnote{Consider a border between 2D Voronoi cells with generators at $(x_1, y_1, z_1=0)$ and $(x_2,y_2,z_2=0)$, with the $y$ metric scaled by a factor $\gamma$ wrt the $x$ metric. That is, the border is the locus of points $(x,y)$ such that $(x-x_1)^2+\gamma^2(y-y_1)^2=(x-x_2)^2+\gamma^2(y-y_2)^2$. One can check that the same border on the $x$-$y$ plane exists in a sectional-Voronoi tessellation, with an isotropic metric and generators at $\left(x_1,\gamma^2y_1,0\right)$ and $\left(x_2,\gamma^2y_2, [(y_2^2-y_1^2)(\gamma^2-\gamma^4)]^{1/2}\right)$. See Fig.\ 6 of \citett{LangBateman2011} for a nice visual depiction of the effect of shearing a spiderweb.} Indeed, it makes intuitive sense that shearing a spiderweb pattern results in something that can still be strung up in tension, albeit likely with different forces. In our case, an anisotropy would show up as a possibly-detectable large-scale gradient in the displacement potential.

\subsection{Example: detecting shear in a Voronoi density field}
\begin{figure}
	\begin{center}
	    \includegraphics[width=0.75\columnwidth]{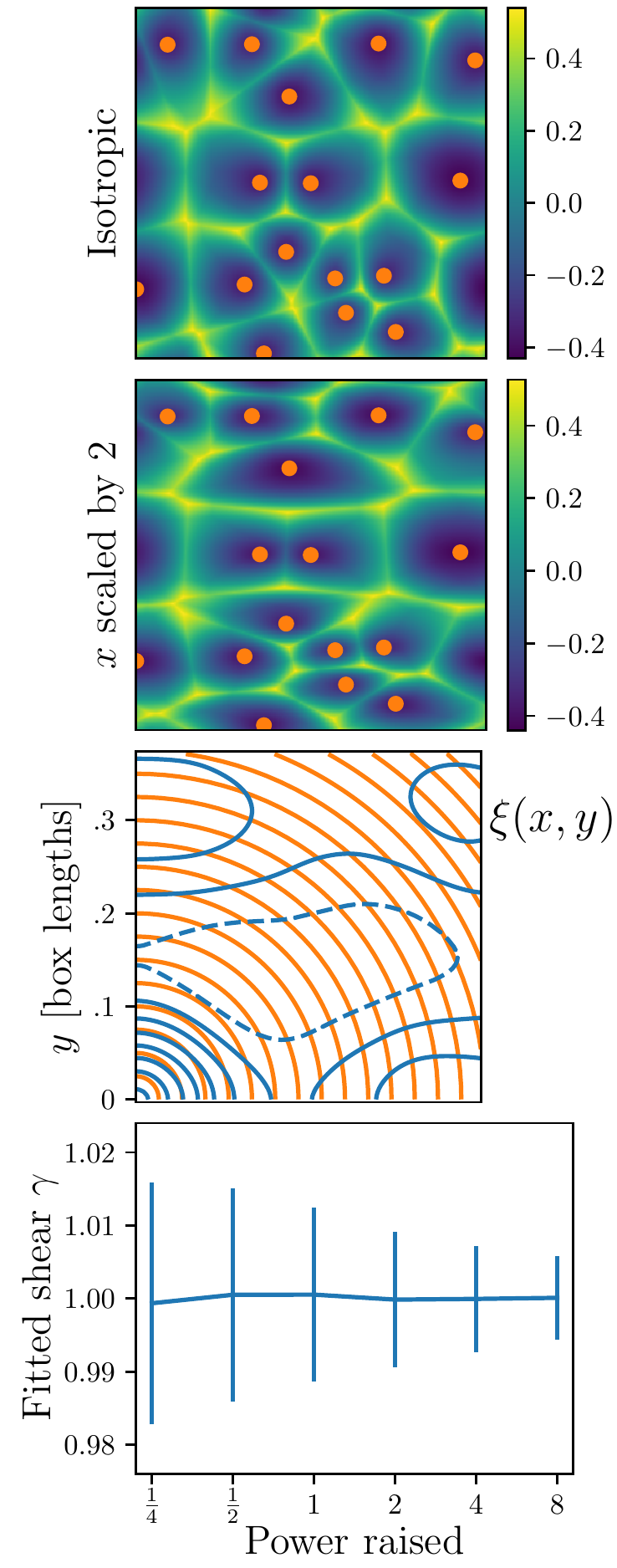}
    	\end{center}
  \caption{Detecting shear in a Voronoi model for a 2D density field. {\it Top panels}: Voronoi density fields from the orange generators, using Eq.\ (\ref{eqn:voronoidensity}), with isotropic ($\gamma=1$) and sheared ($\gamma=2$) distance functions. {\it Third panel}: In blue, contour plot of the 2D two-point correlation function $\xi(x,y)$, measured from the top panel. Orange arcs are contours of constant distance from the origin. The conventional method to measure shear, from $\xi(x,y)$, essentially finds an ellipticity of the orange arcs that best fits the ellipticity of blue contours, with ambiguities in detail. {\it Bottom}: Error bars, typically $\sim 1$\%, in fitting $\gamma$ using the Voronoi shear method described in the text, from a realization such as in the top panel, with input shear $\gamma=1$. Fiducially, `power raised'$=1$, but we explored sensitivity to inaccuracies in assumed void/filament density profiles by raising the `observed' density field to various powers, along the $x$-axis, before fitting $\gamma$.}
  \label{fig:stretchfig}
\end{figure}
Although the sectional-Voronoi situation is more general and accurate in the adhesion model, in the following we turn back to a simplified case of shear in a pure (not sectional) Voronoi tessellation.

Fig.\ \ref{fig:stretchfig} shows the result of shear in a simple 2D Voronoi-based density field, and how well it can be detected. We generated 256$^2$-pixel density fields from sets of 16 Voronoi generators, randomly Poisson-placed except for an exclusion: no two generators are within $b/10$ of each other, where $b$ is the box side length. We used the following model for the density at each point $\bx$:
\begin{equation}
\rho(\bx)=\frac{d_1(\bx)^2}{d_2(\bx) d_3(\bx)}
\label{eqn:voronoidensity}
\end{equation}
where $d_1$, $d_2$ and $d_3$ denote the distances to the first, second, and third nearest Voronoi generators (adding a pixel width to each distance, suppressing staircase-like pixelization effects). As usual in cosmology, we analyze $\delta(\bx)=\rho(\bx)/\bar{\rho}$, with $\bar{\rho}$ the mean density. To our knowledge, this is a new description for a Voronoi-based model for a density field, but resembles that by \citett{Matsubara2007} in using distances to nearest neighbors.

We first tried a density field of the form $d_1(\bx)/d_2(\bx)$, which results in nearly uniform-density filaments, but included the third-nearest generator as well to boost the density at Voronoi vertices over edges, as expected in large-scale structure. In a 3D field, we would include the fourth-nearest neighbor as well, i.e. $\rho=d_1^3/(d_2 d_3 d_4)$, to boost node densities over filament densities. Our model has asymptotically linear profiles for nodes, walls, filaments, and void centers. Perhaps a simple modification or transformation of this prescription could give a good global description of cosmological density profiles away from these morphological features; see \citett{CautunEtal2016} for an example of using wall instead of void density profiles.

To shear the fields, we keep the generators fixed, and change $\gamma$ in the following function giving the distance to a generator position $(x_g,y_g)$, 
\begin{equation}
d_\gamma(x,y)=\sqrt{\frac{1}{\gamma}(x-x_g)^2 + \gamma(y-y_g)^2}.
\end{equation}
In the middle panel, $\gamma=2$. Changing $\gamma$ this much changes the tessellation substantially, as well as noticeably slanting borders between Voronoi cells. The uniform Poisson generator distribution is still isotropic even with a sheared metric, however, except inside the exclusion radius.

There are many tests imaginable to detect shear in a Voronoi or sectional-Voronoi-generated field. Because this section is mainly for illustration, we take a very simplistic approach. We assume that all generator positions are known, and fixed, even when $\gamma$ changes. We fit only for the parameter $\gamma$ in the distance formula, which requires no explicit identification of filaments and nodes of the cosmic web. We generate each realization with $\gamma=1$, and then find the best-fitting $\gamma$ over an ensemble of $\gamma$ near 1. The statistic we maximize over $\gamma$ is $\left\langle\delta_\gamma^{\rm mod}(\bx)\delta_{\gamma=1}^{\rm obs}(\bx)\right\rangle$ i.e.\ the average over pixels of the generated $\delta$ using $\gamma=1$ (the `observed' field) times the same field using a different $\gamma$ (the `model').

In the bottom panel, we show means and standard deviations of the distributions of best-fitting $\gamma$s over 512 realizations, having checked each distribution for Gaussianity. To evaluate sensitivity to shear when the model is imperfect, for the `observed' field, we also raised $\rho(\bx)$ in Eq.\ (\ref{eqn:voronoidensity}) to different powers to obtain $\delta_{\gamma=1}^{\rm obs}$. For $\delta_\gamma^{\rm mod}$, $\rho$ was not raised to a power. In the case that the fitted model was perfect (`power raised'$=1$), there is about a 1\% error in $\gamma$. The error even decreases when the observed field is raised to a high power. This is impressive precision, but it is not so surprising because changing $\gamma$ changes the whole tessellation: not just the orientations of separating walls, but the location of peaks as well. We also checked that adding a visible level of Poisson-sampling noise to the observed field does not substantially degrade error bars.

A standard technique to detect shear in a field is the two-point correlation function $\xi(x,y)\equiv\langle\delta(x^\prime,y^\prime)\delta(x^\prime+x,y^\prime+y)\rangle$; in the third panel, we also show contours of that, measured from the top (isotropic) field. Fitting the isotropy of various contours here would certainly be of use in measuring a shear $\gamma$, but is not obvious what statistic optimally to use, over what range of scales, etc. In a (sheared) Gaussian field, the 2D correlation function would contain all information for doing statistical inference. But this is a highly non-Gaussian field, in which correlation-function information likely falls far short of the full information. We do not attempt a shear measurement from this somewhat noisy correlation function, but we would be surprised if a 1\% measurement could be made from it. We suspect that this Voronoi-based method extracts shear in the field better than the usual correlation function can; in this very idealized example, it may even be optimal.

\section{Conclusion}
In this paper, we have explained the close relationships between a few fields: the large-scale arrangement of matter in the Universe; textile, architectural, and biological spiderwebs; and origami tessellations. The cosmic web forms a spiderweb, i.e.\ a structure that can be strung up to be entirely in tension. If strands of string were strung up in the same arrangement as filaments of a cosmic web, the tension in a string would be proportional to the mass that has collapsed cylindrically onto the filament. This is the cross-sectional area of the boundary between the blobs in the initial Universe that collapsed into the nodes at both ends of the filament.  As far as we know, this is just a geometric correspondence, and the tensions in strands of the structural-engineering spiderweb do not correspond to physical tensions in the cosmos, but perhaps there exists a valid interpretation along these lines.

However, the spiderweb concept rigorously only applies to all nodes (i.e.\ particles, as in an $N$-body simulation), not just those that have experienced collapse. This can be understood in terms of a representation of a cosmic web made from string: to capture kinks and bends in filaments, low-tension strands must be added, likely intersecting outside filamentary regions. Another subtlety is that for the exact correspondence, structure is not considered within dense, multistream regions (haloes, filaments and walls); here, the dark-matter structure is already complicated, and baryonic physics (not treated in the adhesion model) introduces more uncertainties. Thankfully, baryonic effects are expected to leave the structure largely intact outside multistream regions; its complexities are usually activated by shocks that correspond to multistreaming in the dark matter.

It is an open question how far one can relax this simplified problem and retain spiderwebbiness, but successful artistic experiments along these lines suggest that the conditions can be relaxed somewhat. To ease the investigation of this question, we suggest some ways to quantify spiderwebbiness: either measuring the key geometrical property of perpendicularity between nodes of the cosmic web and its dual tessellation, or most engagingly, testing the structural properties of actual tactile representations of it. We also suggest some other ways in which concepts related to spiderwebiness could be used for scientific analysis.

Another question is how much actual webs made by spiders correspond to the cosmic web. Webs by spiders are not always perfect structural-engineering spiderwebs; e.g.\ they can have a strand with slack, indicating a slight departure (perhaps easily fixed by shortening the strand). Even if both arachnid and cosmic spiderwebs were exactly structural-engineering spiderwebs, likely they would differ in some quantifiable properties, e.g.\ in the distribution of the number of strands coming off a node.

Geometry (specifically here, the concept of a Voronoi-related tessellation) provides the link between these apparently disparate scientific and artistic fields, as it does for many others \citepp[e.g.][]{Kappraff2001,Senechal2013}. If a spider did construct the cosmic web (an idea we do not advocate), that spider was not ignorant of geometry.







\section*{Data Access}
See \hrefurl{https://github.com/jhidding/adhesion-example/blob/master/notebook/Adhesion.ipynb} for a Python notebook, with many figures illustrating the adhesion model, visible within a browser. This software, written by JH and citable as \citepp{Hidding2017zenodo}, emanates from NWO project 614.000.908 supervised by Gert Vegter and RvdW. In that repository, \href{https://github.com/jhidding/adhesion-example/blob/master/notebook/Adhesion_spiderweb_origami.ipynb}{Adhesion\_spiderweb\_origami.ipynb} generates some figures in the paper. Also see an interactive sectional-Voronoi spiderweb demonstration and designer at \hrefurl{https://github.com/neyrinck/sectional-tess}, usable without installation at \hrefurl{https://mybinder.org/v2/gh/neyrinck/sectional-tess/master} (run notebook sec\textunderscore voronoi\textunderscore spiderweb.ipynb).

\section*{Authors' Contributions}
MN conceived, undertook, and wrote up the current project. JH provided essential clarifying adhesion-model code that enabled many of the figures, and explanations of the sectional-Voronoi formalism that he developed. MK shared substantial structural-engineering expertise and guidance, and clarified descriptions in the text. RvdW is leading the Groningen cosmic web program, major components of which are the adhesion model project, and additional tessellation-related projects. He has also contributed clarifications to the text and on tessellation-related issues.

\section*{Funding}
MN thanks the UK Science and Technology Facilities Council (ST/L00075X/1) for financial support. JH and RvdW acknowledge support by NWO grant 614.000.908.

\section*{Acknowledements}
MN is grateful for encouragement and discussions about this work at the scientific/artistic interface from Fiona Crisp, Chris Dorsett, and Si\'{a}n Bowen at the Paper Studio Northumbria; Sasha Englemann and Jol Thomson; Alex Carr, Giles Gasper and Richard Bower at Durham, of the Ordered Universe Project; origamist Robert Lang; and SciArt Center, where MN was a 2016-2017 `The Bridge: Experiments in Science and Art' resident, with Lizzy Storm. MN thanks Hannah Blevins for expertise, advice and encouragement for the textile spiderweb in Fig.\ \ref{fig:councilofgiants}; and Reece Stockport and Ryan Ellison for 3D printing help leading to Fig.\ \ref{fig:threedprint}. JH and RvdW wish to thank Gert Vegter, Bernard Jones and Monique Teillaud for helpful discussions about the adhesion formalism, Monique Teillaud in particular for her help and support with CGAL. We also thank Allan McRobie (particularly MK, also for PhD supervision), Masoud Akbarzadeh, and Walter Whiteley for useful discussions.

\bibliographystyle{mnras}
\bibliography{refs}

\end{document}